\documentclass[Times,12pt,oneside,openany,print,index]{report}
\usepackage[a4paper,width=150mm,top=25mm,bottom=25mm]{geometry}
\usepackage[english]{babel}
\usepackage[utf8]{inputenc}
\usepackage{csquotes} 
\usepackage{amsmath} 
\usepackage{gensymb}
\usepackage{listings}
\usepackage{color}
\usepackage{listings}
\usepackage{color}
\usepackage[LGR,T1]{fontenc}
\usepackage[utf8]{inputenc}   

\newcommand{\textgreek}[1]{\begingroup\fontencoding{LGR}\selectfont#1\endgroup}

\definecolor{dkgreen}{rgb}{0,0.6,0}
\definecolor{gray}{rgb}{0.5,0.5,0.5}
\definecolor{mauve}{rgb}{0.58,0,0.82}

\definecolor{dkgreen}{rgb}{0,0.6,0}
\definecolor{gray}{rgb}{0.5,0.5,0.5}
\definecolor{mauve}{rgb}{0.58,0,0.82}

\usepackage{graphicx} 
\graphicspath{/images} 
\usepackage{caption} 

\usepackage{array} 

\usepackage[nottoc]{tocbibind} 

\usepackage[normalem]{ulem} 

\usepackage{hyperref} 
\hypersetup{
    colorlinks=true,
    linkcolor=black,
    filecolor=magenta,      
    urlcolor=blue,
    citecolor=blue,
}
\usepackage{nomencl} 
\renewcommand{\nompreamble}{The following list defines several acronyms \& abbreviations that are used in the body of the document.}
\makenomenclature

\usepackage[backend=biber,style=apa,sorting=ynt]{biblatex}

\usepackage{amssymb}

\let\cleardoublepage=\clearpage 

\begin{document}

\thispagestyle{empty} 
\begin{titlepage}
\renewcommand*{\thepage}{Title} 

    \begin{center} 
        \vspace*{3cm} 
        
        {\fontsize{24pt}{22pt}\selectfont{Study and Design of\\Reconfigurable Intelligent Surfaces\\
        \ \\
        \ \\
        \ \\
        \fontsize{22pt}{20pt}\selectfont{\textgreek{Μελέτη και Σχεδίαση\\Επαναπροσδιοριζόμενων Ευφυών Επιφανειών}}}
        } 
        
        \vspace{1.5cm}
        
        \text{}
        
        \vspace{0.5cm}
        
        	{\fontsize{16pt}{22pt}\selectfont{Apostolos Spanakis-Misirlis}}\\
	        {\fontsize{14pt}{22pt}\selectfont{$\Pi$18183}}\\

        \ \\
        \ \\
        \ \\
        \ \\

         Supervisor: Christos Douligeris

        \vspace{3cm}
        
        	A thesis submitted to the Department of Informatics\\
            in partial fulfillment of the requirements for the degree of\\
            B.Sc. in Informatics

        \vspace{2.5cm}
        
    		School of Information and Communication Technologies\\
            University of Piraeus\\
            September 2022
        
        \vspace{3cm}
        
    
    \end{center}

\end{titlepage} 
\cleardoublepage

\pagenumbering{roman} 




\phantomsection
\addcontentsline{toc}{chapter}{Abstract}
\ \\
\ \\
\section*{Abstract}
In this thesis, we introduce the fundamental equations behind the estimation of the link budget in a communications channel, highlighting the key limitations of conventional systems. Furthermore, we investigate the use of reconfigurable intelligent surfaces as a modern method of overcoming obstruction losses, while making use of numerical methods and computational electromagnetics to understand its physical mechanism and probe its theory of operation. Additionally, a preprint on computational geometry is presented, applicable to the field of computational electromagnetics, enabling the simulation of systems such as reconfigurable intelligent surfaces using open-source tools. Lastly, we provide a tool for the physical optimization of radio-frequency networks, based on mathematical programming. Such a tool may be used for the optimization of reconfigurable intelligent surfaces, ultimately improving the communication channel between a transmitter and receiver.\\
\ \\
\ \\
\ \\
\ \\

\section*{\textgreek{Περίληψη}}

\textgreek{Σε αυτή την πτυχιακή, εισάγουμε τις θεμελιώδεις εξισώσεις πίσω από την εκτίμηση του ισοζυγίου ζεύξης σε ένα κανάλι επικοινωνίας, επισημαίνοντας τους βασικούς περιορισμούς των κλασικών συστημάτων. Επιπλέον, διερευνούμε τη χρήση επαναπροσδιοριζόμενων ευφυών επιφανειών ως σύγχρονη μέθοδο αντιμετώπισης απωλειών διάδωσης, χρησιμοποιόντας αριθμητικές μεθόδους και υπολογιστικό ηλεκτρομαγνητισμό για την κατανόηση του φυσικού μηχανισμού και τη διερεύνηση της θεωρίας πίσως από την λειτουργία τους. Επιπλέον, παρουσιάζεται ένα άρθρο σχετικά με την υπολογιστική γεωμετρία, που εφαρμόζεται στον τομέα του υπολογιστικού ηλεκτρομαγνητισμού, επιτρέποντας την προσομοίωση συστημάτων όπως είναι οι επαναπροσδιοριζόμενες ευφυές επιφάνειες, χρησιμοποιώντας εργαλεία ανοικτού κώδικα. Τέλος, παρέχουμε ένα εργαλείο για τη φυσική βελτιστοποίηση των δικτύων ραδιοσυχνοτήτων} (RF networks), \textgreek{το οποίο βασίζεται στον μαθηματικό προγραμματισμό. Ένα τέτοιο εργαλείο μπορεί να χρησιμοποιηθεί για τη βελτιστοποίηση επαναπροσδιοριζόμενων ευφυών επιφανειών, βελτιώνοντας τελικά τον δίαυλο επικοινωνίας μεταξύ πομπού και δέκτη.}

\vspace{1cm}
\pagebreak

\phantomsection
\addcontentsline{toc}{chapter}{Dedication}
\section*{Dedication}
\vspace{15pt}
\textit{To those who value open science.}

\pagebreak

\phantomsection
\addcontentsline{toc}{chapter}{Acknowledgment}
\section*{Acknowledgements}
\vspace{15pt}
I wish to thank Dimitrios Karagiannis, Cameron Van Eck, and Athanasios Kanatas for generously taking the time to provide their valuable teachings in mathematics, physics, and engineering, respectively.

\renewcommand{\contentsname}{Table of Contents} 
\cleardoublepage
\phantomsection
\addcontentsline{toc}{chapter}{Table of Contents} 
\tableofcontents 

\listoffigures 
\listoftables 

\printnomenclature 
\addcontentsline{toc}{chapter}{Nomenclature}
\cleardoublepage

\pagenumbering{arabic} 
\setcounter{page}{11}

\chapter{Introduction}
\section{Challenges in Telecommunications }
For the last few decades, the field of telecommunications has enabled a vast range of applications, ranging from radio and television, to the relatively recent support of the internet, mobile communications, and others. Although the applications this area of technology has introduced are undoubtedly significant, it is worth noting that these implementations do not always come without major technical challenges. When designing a wireless system to take place in a communication channel consisting of a transmitter and a receiver, several parameters must be considered in order to evaluate the feasibility of the link. Unless each and every one of these parameters is taken into account, it is especially difficult to accurately estimate the reliability of a wireless link.\\

The thesis is structured as follows: in this Chapter, we introduce the fundamentals behind the calculation of link budgets used to evaluate the reliability and quality of a communications channel. In Chapter 2, we cover the theory of reconfigurable intelligent surfaces as a way of overcoming obstruction losses, and highlight the most common methods of operation. Chapter 3 deals with the use of computational electromagnetics to tackle the problem of understanding the theory of operation of intelligent surfaces at the meta-atom level, while Chapter 4 introduces an algorithm to deal with geometry discretization for electromagnetic simulations; similar to those carried out in Chapter 4. Finally, in Chapter 5 we present an open-source modular optimization interface for network analyzers, which can be used to optimize the transfer function of filters, reconfigurable intelligent surfaces, and other RF networks.

\section{Link Budget Evaluations }
The computation of the so-called link budget, describing the total power lost from a transmitter to a receiver, is very straightforward to evaluate. Given all the necessary parameters are known with little error, we can estimate the properties of a link with great accuracy, using the simplified equation

\begin{align} \label{eq:1.1}
    P_{\mathrm{RX}} = P_{\mathrm{TX}} + \mathrm{Gains} - \mathrm{Losses},
\end{align}

where $P_{\mathrm{RX}}$ and $P_{\mathrm{TX}}$ are the received and transmitted power in dBm, respectively. Note that it is generally useful to express all gain and loss quantities in decibels (dB), because due to the properties of logarithms, we can simply add and subtract (instead of multiplying and dividing), which is algebraically easier to compute.

\subsection{Parameters Influencing Wireless Links }
Although Equation (\ref{eq:1.1}) is easily interpretable, identifying the exact values of the gains and losses requires the investigation of their underlying components.\\

Specifically, the gains of the link can be divided into the following sub-parameters:
\begin{itemize}
    \item TX antenna gain, $G_{\mathrm{TX}}$ (dBi)
    \item RX antenna gain, $G_{\mathrm{RX}}$ (dBi)
\end{itemize}

These quantities are often combined and referred to by the effective isotropic radiated power (EIRP), describing the total power radiated in the direction of the main lobe of the transmitting antenna (boresight),

\begin{align} \label{eq:1.2}
    \mathrm{EIRP}_\mathrm{dB} = P_{\mathrm{TX}} - L_{\mathrm{TX}} + G_{\mathrm{TX}},
\end{align}

Oppositely, link losses can be split into:
\begin{itemize}
    \item TX losses, $L_{\mathrm{TX}}$ (dB)
    \item Free-space path loss, $L_{\mathrm{FS}}$ (dB)
    \item Other miscellaneous losses (e.g. intervening obstacles), $L_{\mathrm{m}}$ (dB)
\end{itemize}

\begin{figure}[htbp]
\centering
\includegraphics[scale=0.8]{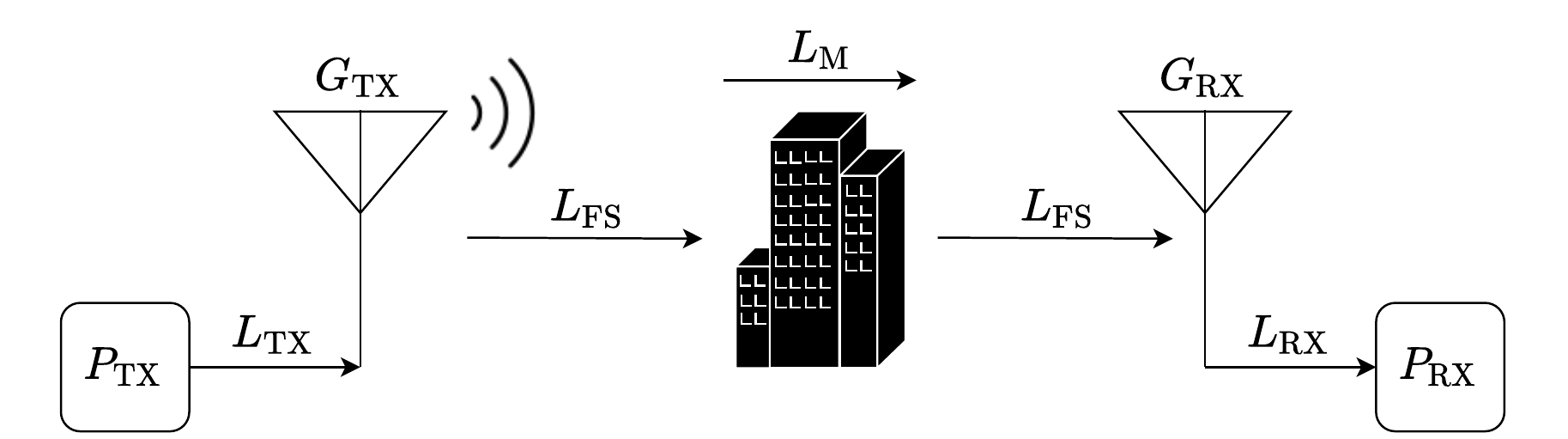}
\caption{A flow diagram depicting the gains and losses of a wireless link, all of which are taken into consideration during the computation of the link budget.}
\label{fig:1.1}
\end{figure}

A basic example of a communication channel involving the gains and losses of the link is given in Fig.~\ref{fig:1.1}. Note that, despite the (usually) minimal losses introduced by the transmitting and receiving systems via $L_{\mathrm{TX}}$ and $L_{\mathrm{RX}}$, respectively, the primary source of signal attenuation is generally a form of path loss.

\begin{figure}[htbp]
\centering
\includegraphics[scale=0.8]{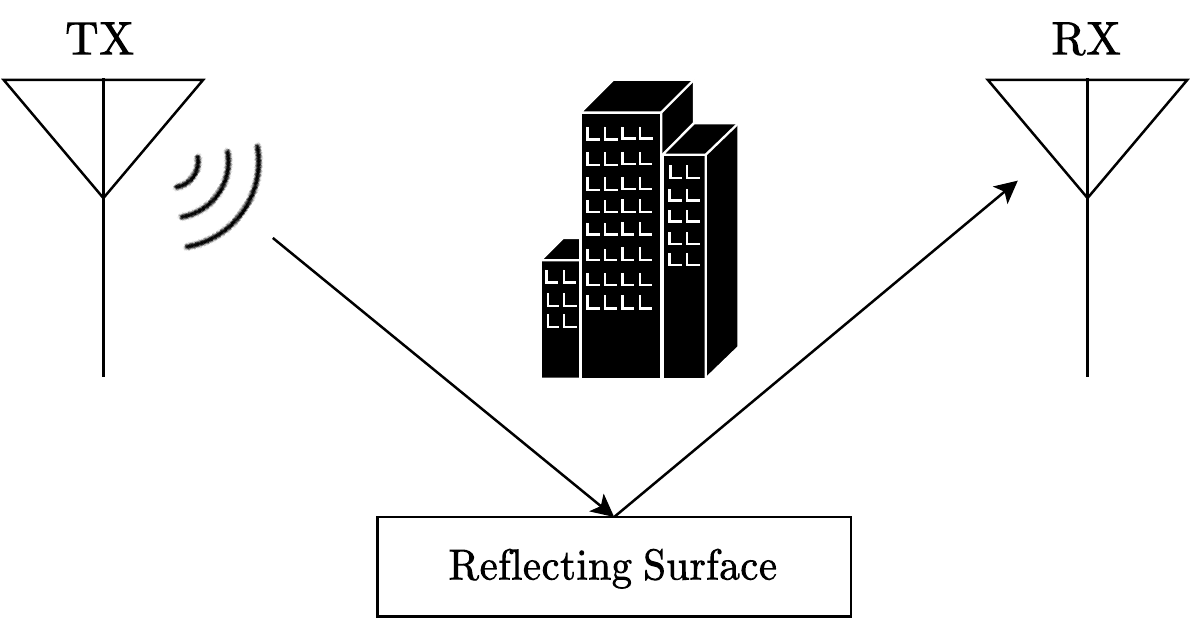}
\caption{A reflecting surface placed between the transmitter and receiver, enabling a new communication channel around intervening obstacles.}
\label{fig:1.2}
\end{figure}

\subsection{Free-Space Path Loss}

The free-space path loss, $L_{\mathrm{FS}}$ refers to the line-of-sight attenuation between two communication nodes, and like any other form of path loss, it is independent of the transmitting and receiving systems. Following the inverse-square law stating that intensity is directly proportional to the square of the distance, this loss is given by

\begin{align} \label{eq:1.3}
    L_{\mathrm{FS}} = 10\log_{10} \left[ \left(\frac{4 \pi d}{\lambda}\right)^2 \right],
\end{align}

where $d$ is the distance between two communication nodes and $\lambda$ is the wavelength of the electromagnetic wave being transmitted across the wireless channel. As it is quite common to work in terms of frequency, $f$, Equation (\ref{eq:1.3})  may also be rewritten as

\newcommand{\eqdef}{\overset{f = \frac{c}{\lambda}}{=\joinrel=\joinrel=}}

\begin{align} \label{eq:1.4}
    L_{\mathrm{FS}} \eqdef 20\log_{10} \left(\frac{4 \pi d f}{c}\right),
\end{align}

where $c$ is the speed of light in the medium of propagation. It may appear obvious that at realistic distances and common radio frequency bands, the free-space path loss can have a significant impact on the received power, degrading the signal-to-noise ratio to a considerable degree.

\subsection{Obstruction Losses}

Besides the free-space path loss, there are also other types of miscellaneous losses that are not directly part of the transmitter or receiving system. One loss that can have a major impact on the quality of a communication link is the attenuation introduced by physical obstacles. If such intervening barriers are considerably large and located at inauspicious positions between the transmitting and receiving antennas, $P_{\mathrm{RX}}$ may very well sink down to an unacceptable level. Of course, such obstacles are dependent on the physical environment the wireless channel takes place in, and due to a variety of intervening objects and natural landscapes, these losses are commonly identified in both indoor and outdoor channels.\\

One potential solution to this problem is the introduction of a reflecting surface between the two communication nodes. Placed in a particular position and orientation, such that the radiated electromagnetic waves get reflected toward the direction of the receiving end (without being reflected or absorbed by intervening structures), the improvement of $P_{\mathrm{RX}}$ becomes more feasible. A simple version of this configuration is shown in Fig.~\ref{fig:1.2}, and it is obvious that due to its reciprocal properties, a bidirectional (duplex) communication channel could be supported just as well.\\

However, the simplicity of this configuration comes with a major pitfall: the operation of such a channel assumes both antennas are perfectly stationary and located in an ideal direction relative to the orientation of the reflecting surface. In many real-world telecommunication scenarios though, such assumptions are invalid, especially in cellular networks and mobile communications.

\nomenclature{dB}{Decibel}
\nomenclature{dBm}{Decibel milliwatts}
\nomenclature{dBi}{Decibel isotropic}
\nomenclature{TX}{Transmitter}
\nomenclature{RX}{Receiver}
\nomenclature{EIRP}{Effective isotropic radiated power}

\chapter{Reconfigurable Intelligent Surfaces}
\section{Theory of Operation}
In the previous chapter, the effects of introducing a reflecting surface between two antennas were discussed. As mentioned, the evident drawback of such a technique is the requirement of the transmitting and receiving ends to be located at just the right position in space, relative to the orientation and placement of the intervening reflecting surface. If this condition is not fulfilled, little to no electromagnetic radiation will manage to reach the receiver.\\

For the past few years, a solution that has been proposed to tackle this problem is the use of a so-called reconfigurable intelligent surface (RIS). In contrast to traditional simple reflecting surfaces, RISs have the capability of electronically adjusting their electromagnetic properties (generally in a programmable manner), with the goal of controlling and optimizing the propagation of electromagnetic radiation to the advantage of the wireless link.

\subsection{Snell's Law}
According to Snell's law, a traditional reflecting surface behaves in a very straightforward manner. If we assume a common medium with a constant refractive index ($n_1 = n_2$), the incident ray's angle, $\theta_i$ will always be equal to the reflected ray's angle, $\theta_r$. This is easily derived from the generalized ratio of sines:

\begin{align} \label{eq:2.1}
    \frac{\sin \theta_r}{\sin \theta_i} = \frac{n_1}{n_2},
\end{align}

which can be algebraically transformed to prove that

\begin{align} \label{eq:2.2}
    \frac{\sin \theta_r}{\sin \theta_i} = 1\\
    \therefore \ \ \theta_i = \theta_r \ \  \label{eq:2.3}
\end{align}

\begin{figure}[htbp]
\centering
\includegraphics[scale=1.7]{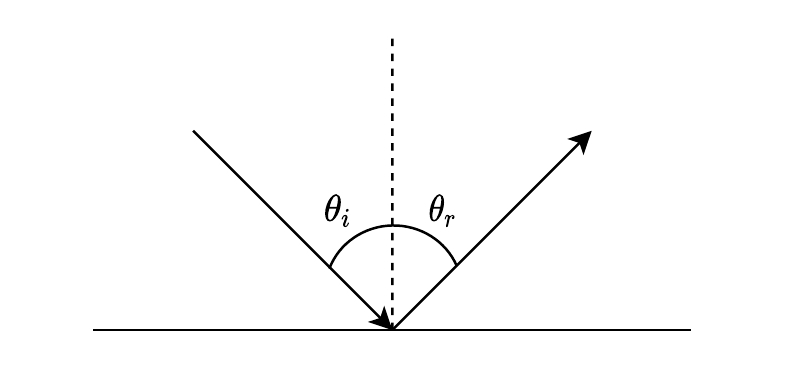}
\caption{Visual representation of Snell's law, demonstrating the clear relationship between angles $\theta_i$ and $\theta_r$.}
\label{fig:2.1}
\end{figure}

This relationship is visually presented in Fig.~\ref{fig:2.1}, and it applies across the entire electromagnetic spectrum: from radio and microwaves, to visible light and beyond.

When designing a reconfigurable intelligent surface, the ultimate goal is to artificially `violate' or `get around' this restriction imposed by Snell's law, using certain clever approaches.

\subsection{Diffuse Reflection}
\begin{figure}[htbp]
\centering
\includegraphics[scale=0.5]{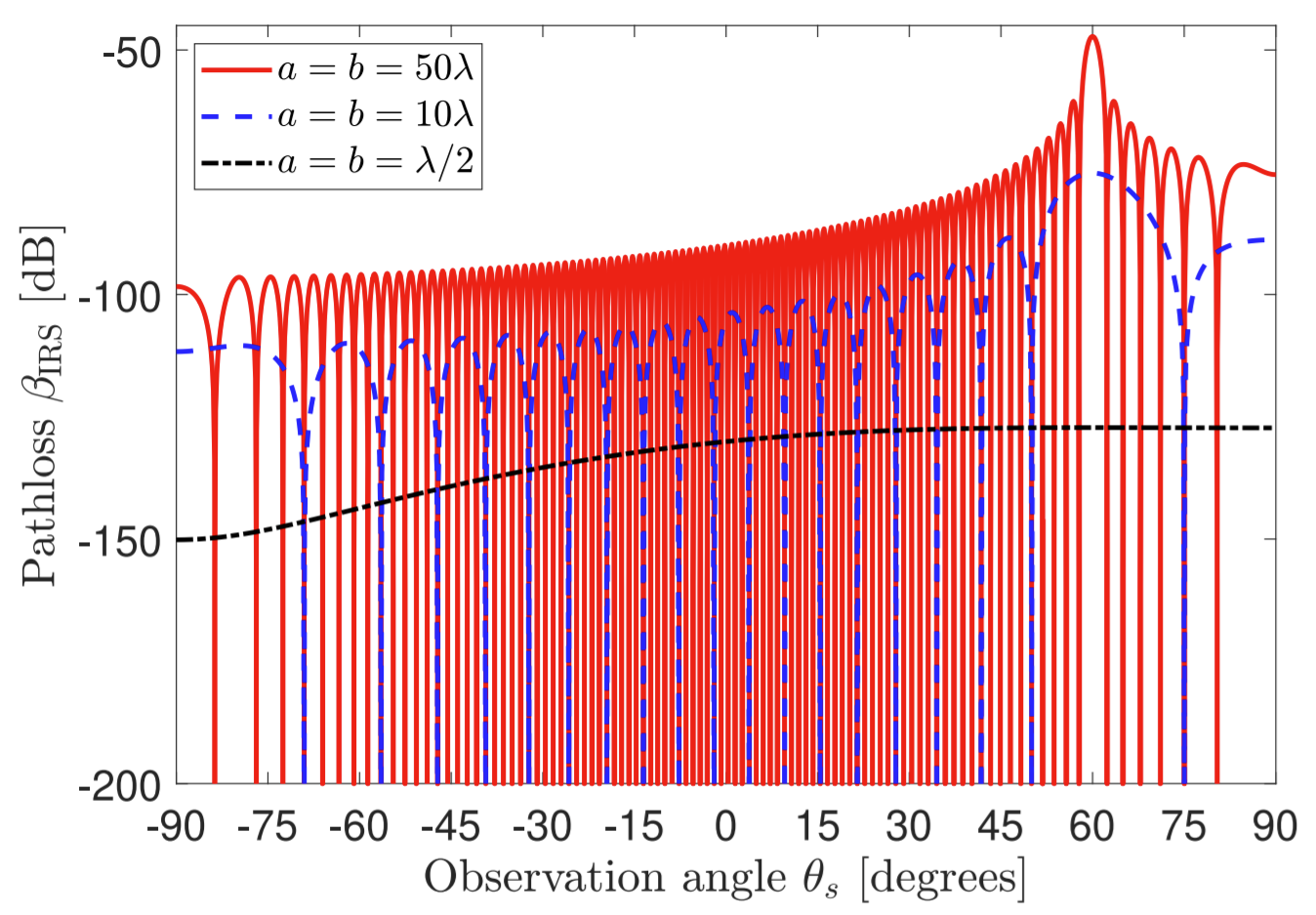}
\caption{The pathloss from the reflecting surface to the receiver, for a particular configuration ($G_{\mathrm{TX}}=G_{\mathrm{RX}}=5\ \mathrm{dBi}$). The distance between the point source and the surface is 50 m, while the reflected path's distance is 25 m. \copyright \ 2019 IEEE}
\label{fig:2.2}
\end{figure}

The limits of Snell's law Equation (\ref{eq:2.3}) start to become particularly apparent when the reflecting surface approaches sub-wavelength dimensions. At such small scales, the incident wave begins to no longer follow specular reflection ($\theta_i = \theta_r$), but is instead scattered in multiple directions around $\theta_r$. This optical effect is known as diffuse reflection, and thanks to its properties, the operation of intelligent reflecting surfaces is enabled. Combined in an appropriate configuration, these sub-wavelength meta-atoms can ultimately initiate wave interference (constructive and destructive), supporting the formation of beams towards particular directions of interest. Fig.~\ref{fig:2.2} presents analysis (\cite{Ozgecan2020}) demonstrating the non-specular optical response of a surface to an incident plane wave at $\theta_i=30\degree$ and a desired direction of $\theta_r=60\degree$.

\subsection{Beam Reconfigurability}
\begin{figure}[htbp]
\centering
\includegraphics[scale=1]{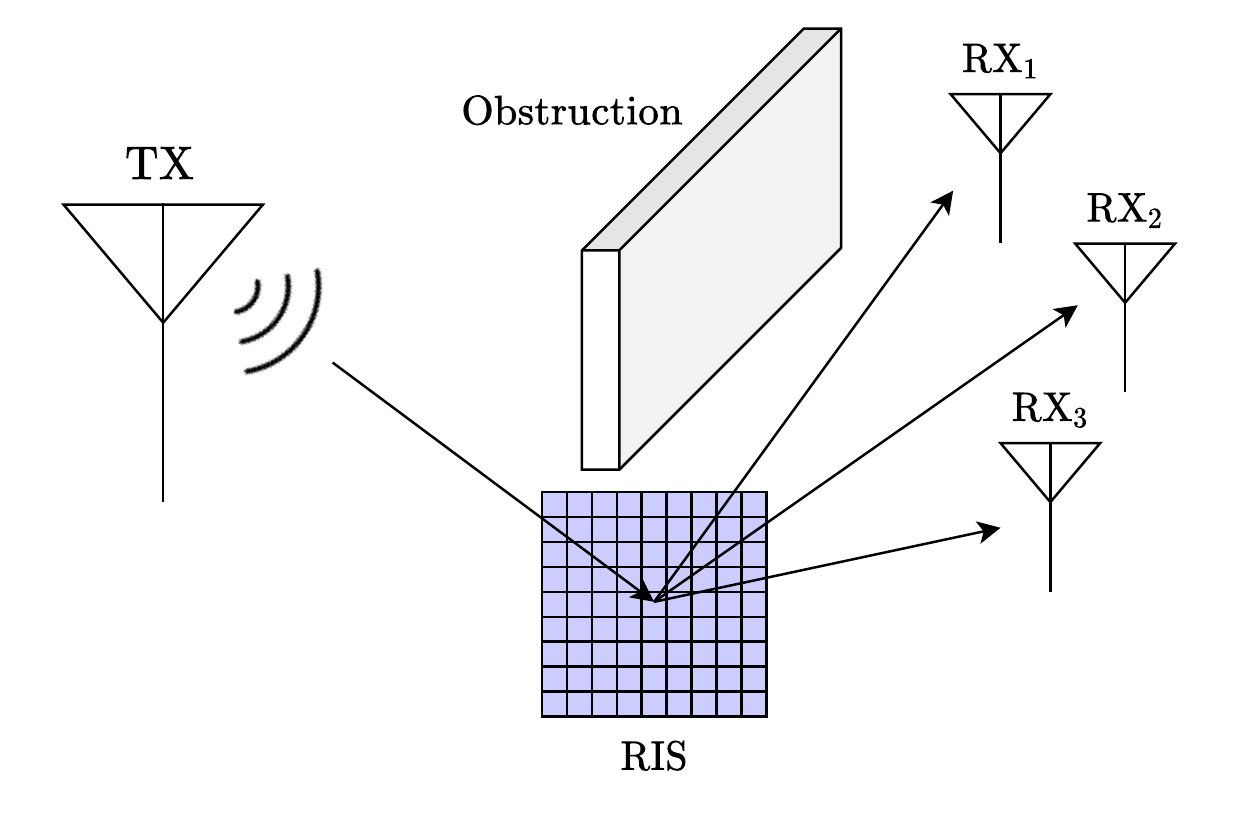}
\caption{A reconfigurable intelligent surface installed between a transmitter and three receivers, supporting three wireless links around the intervening obstacle.}
\label{fig:2.3}
\end{figure}

The main idea behind reconfigurable intelligent surfaces is the ability to programmatically adjust their behavior to incident electromagnetic waves. The most common implication of this is beam reconfigurability. Being able to adjust the angle $\theta_r$ for any given $\theta_i$ is a major luxury, as it can uncover new channels of communication for multiple new users, regardless of their direction\footnote{Generally true as long as each receiver is not too far off the surface's angle of coverage.} relative to the RIS. This idea is demonstrated in Fig.~\ref{fig:2.3}, where three receivers are supported using a single intelligent reflecting surface.\\

Note that the number of supported receiving clients can in theory be arbitrarily large, depending on the capabilities of the RIS system. Even if the formation of multiple individual beams is limited to a small number of users simultaneously, clever techniques (such as rapid beamsteering based on individual user activity) can still be employed to maximize coverage for as many users as possible.

\section{Challenges and Limitations}
Since the proposal of the use of intelligent reflective surfaces in present and future telecommunication applications is a very recent concept with many open questions, the research and investigation of the underlying issues and restrictions surrounding this novel technology is of utmost importance. Furthermore, in order to classify this new method as advantageous, we have to compare it with other conventional techniques used in today's world, and identify the key fields in which RISs can play a game-changing role in enabling new ways of transferring information efficiently. To achieve this, three major aspects must be analyzed and taken into consideration: performance, size and cost.

\subsection{Performance}
Although an accurate estimation of the performance of an intelligent reflecting surface is a difficult task without the use of electromagnetic simulations, we can still compute an approximate estimate end-to-end channel gain from the transmitter to the receiver.\\

Let $g_n$ be the channel from the transmitter to a meta-atom $n$ of the RIS.\footnote{To simplify calculations, we shall neglect any RIS-induced losses, which are often minimal.} Due to the properties of the RIS, a phase shift of $e^{j\phi_n}$ will be introduced. If we denote the receiving-path channel (from the RIS to the receiver) as $h_n$, the end-to-end channel gain will simply be the product of

\begin{align} \label{eq:2.4}
    k_n = g_n e^{j\phi_n} h_n.
\end{align}

If we extend the total number of atoms the RIS consists of to $m$, the received signal takes the form of

\begin{align}
    \displaystyle\sum_{n=1} ^{m} k_n \cdot S+N = \sum_{n=1} ^{m} g_n e^{j \phi_n} h_n \cdot S+N,
\end{align}

where $S$ and $N$ correspond to the signal and noise, respectively. In order to compute the end-to-end channel gain, Equation (\ref{eq:1.3}) can be used to derive the free-space path loss from the transmitter to the RIS ($d_g$) and from the RIS to the receiver ($d_h$). If we denote the aperture of an individual meta-atom as $A$, then for any distance $d$, the channel gain from/to that atom becomes

\begin{align}
    \frac{A}{4\pi d^2}.
\end{align}

The end-to-end channel gain is therefore

\begin{align}
    |g_n|^2 |h_n|^2 = \frac{A}{4\pi d_g} \cdot \frac{A}{4\pi d_h},
\end{align}

and since the phase shift cancels out while taking the square of the magnitude of $k_n$ (Equation \ref{eq:2.4}), the  channel gain from the transmitter to the receiver is

\begin{align} \label{eq:2.8}
    \frac{A^2}{{(4 \pi d_g d_h)}^2}.
\end{align}

Finally, to further extend this to $m$ elements (taking the entirety of the intelligent reflecting surface into account), we can multiply Equation (\ref{eq:2.8}) by $m^2$ (accounting for both the incident and the return path), ending with

\begin{align} \label{eq:2.9}
    \frac{m^2 A^2}{{(4 \pi d_g d_h)}^2}.
\end{align}

At this point, it is worth emphasizing that despite the appealing simplicity of this equation, accurately estimating the channel of a wireless system involving intelligent surfaces is far from straightforward. This stems from the fact that Equation (\ref{eq:2.9}) assumes all $m$ meta-atoms exhibit an electromagnetically identical behavior, and the nonuniformity of certain regions across the surface is not taken into consideration.\\

The estimation of an intelligent surface's performance also gets increasingly complicated to compute when multiple beams are formed (e.g. to cover multiple users), or when direction-dependent radio-frequency interference (RFI) is present. In fact, one of the open problems surrounding RISs is the suppression of RFI and the identification of the most suitable configuration for the optimal phase shift for each individual meta-atom ($\phi_n$), with the goal of ultimately maximizing the end-to-end channel gain for a particular receiver.

\subsection{Size and Cost}
One of the most promising applications of reconfigurable intelligent surfaces is mobile communications. The main speculation that is commonly discussed is that in certain telecommunication scenarios, intelligent surfaces have the potential to replace base stations. This idea originates from the fact that base stations often tend to have a very high installation and operation cost (due to size and electric power consumption costs), and can even take a long time and effort to set up in a particular location. The expense of such an operation may become exceptionally large (relative to the potential profit it would offer) when the coverage of interest is merely a small isolated region.\\

On the other hand, if RISs are instead to be deployed, due to their (usually) passive nature and physical simplicity, they have a promising potential to replace base stations with little to no sacrifice in the performance of the link. In fact, at particularly high frequencies (short wavelengths), reconfigurable intelligent surfaces can get even smaller. This is due to the fact that the dimensions of a RIS obviously scale with the wavelength of operation. Despite the capabilities of passive RISs, active types of RISs have also been proposed (\cite{zhang2021active,zhi2021active}), some of which are targeted to future sixth-generation ($6\mathrm{G}$) networks, which operate by amplifying the incident signal using low-noise amplifiers (LNAs). This technique could further improve the link budget for a system involving a RIS.

\section{Types of RIS}

Reconfigurable intelligent surfaces are based on the reconfigurability property of the atoms (elements) they consist of. By selectively adjusting the impedance of each individual element, we are ultimately able to simulate a different kind of surface, without having to physically alter its geometric structure. In terms of hardware architecture, the design of reconfigurable intelligent surfaces can take many forms to achieve impedance reconfigurability. In this section, we will explore the two main types of techniques to achieve this: namely, based on the operation of PIN diodes, and varactor diodes.

\subsection{PIN Diodes}

\begin{figure}[htbp]
\centering
\includegraphics[scale=1.9]{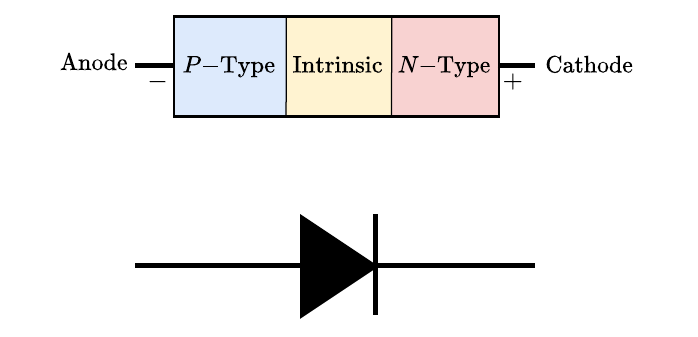}
\caption{The structure of a PIN diode, highlighting the position of each layer (top), along with its electronic symbol used in schematic circuits (bottom). The orientation of the triangular shape denotes the position of the diode's anode (A) and cathode (K) terminals.}
\label{fig:2.4}
\end{figure}

PIN diodes are diodes consisting of a wide and undoped intrinsic semiconductor layer, surrounded by a P-Type and an N-Type semiconductor. Unlike the intrinsic (I-type) semiconductor, the P-type and N-type regions are generally doped heavily. The width of the intrinsic region allows RF networks like attenuators and switches to be designed in a simple, compact, and low-cost manner, using inexpensive PIN diodes. Fig.~\ref{fig:2.4} depicts the structure of a PIN diode, along with the diode's electronic symbol.

Given the popularity of PIN diodes in recent years, it is common practice to make use of this kind of electronic component in a variety of RF devices, including antennas. This is because PIN diodes offer a very inexpensive and straightforward interface to toggle various configurations as binary-type switches (ON/OFF) on a network, controllable using a simple voltage source. For that reason, their implementation has been very popular in, e.g., reconfigurable antenna designs, where the radiation pattern, operating frequency range, or polarization characteristics demand reconfigurability. Likewise, PIN diodes can be just as useful in the design of intelligent reconfigurable surfaces.\\

The way this works is similar to the operation of reconfigurable antennas. Instead of having a simple patch element (or atom, in the case of RIS), we integrate a PIN diode on the element in a particular configuration, such that the ON/OFF toggling of the diode introduces a shift in the impedance. This implies that each element of the intelligent surface will exhibit a different phase response, depending on the state of its associated diodes.

\subsection{Varactors}

Another component that enables reconfigurability in RF networks is a varactor. Just like PIN diodes, varactors have also been increasingly popular in the field of RF engineering, and their implementation can be found in reconfigurable antennas too. There are two main differences between varactor and PIN diodes: firstly, varactors alter the capacitance based on the bias voltage, while PIN diodes are limited to two binary-state options:

\begin{align} \label{eq:short_open}
    R=\mathrm{Re}(Z)&\approx 0 \mathrm{\ \ \  \ \ \ \ \ (short),\ and}\\
    R&\approx \infty \mathrm{\ \ \ \ \ \ \ (open).  \ \ \ \ \ \ \ \ \ \  \ \  }
\end{align}

This makes varactor diodes a lot more flexible, allowing the controller to alter between multiple states, with richer versatility.\\

\begin{figure}[htbp]
\centering
\includegraphics[width=\textwidth]{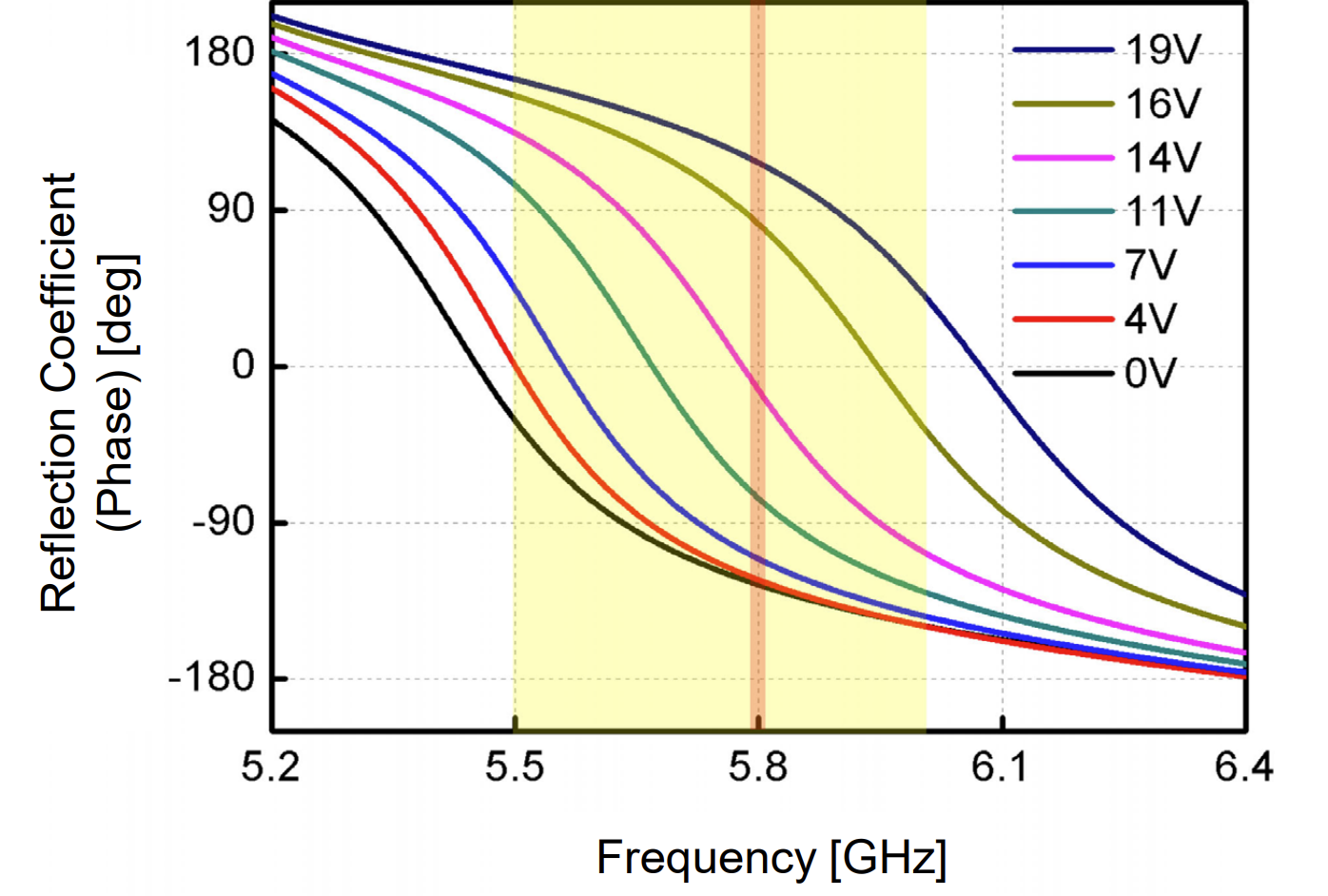}
\caption{Phase response as a function of bias voltage and frequency. Notice that at the frequency of interest (5.8 GHz), a wide range of phase responses can be obtained by varying the varactor bias voltage. The yellow region highlights the frequency range across which the maximum phase difference exceeds $180^{\circ}$.}
\label{fig:phase_response}
\end{figure}

\begin{figure}[htbp]
\centering
\includegraphics[width=\textwidth]{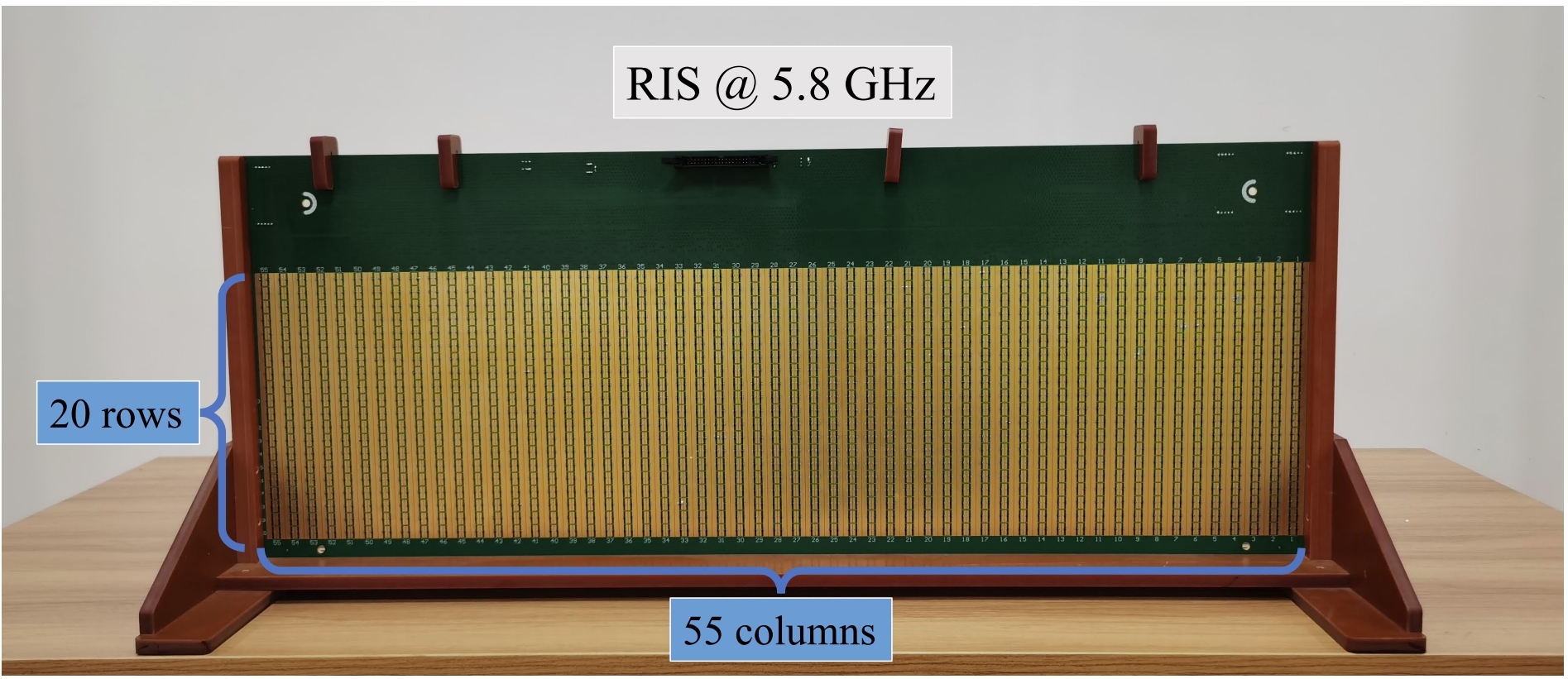}
\caption{Intelligent surface prototype consisting of varactor-based elements, designed for operation at 5.8 GHz. Note that the length and width of each atom are less than 15 mm. At a wavelength of 5.17 cm, this corresponds to a length of $\lambda/3.45$. The reason for the necessity of such small scales in the design of RIS atoms is explored in the following chapter.}
\label{fig:ris}
\end{figure}

Similar to PIN diodes, the role of varactor diodes in intelligent surfaces is to alter the impedance of their associated atom, yielding a different phase response. Fig.~\ref{fig:phase_response} highlights the effect on the phase response of an atom, introduced by the integration of a pair of varactor diodes, as a function of frequency. This configuration was used to build a large RIS surface consisting of 1100 atom elements (Fig.~\ref{fig:ris}) as a proof of concept, which served as a successful experiment to showcase the communication improvement introduced by the presence of the RIS between the transmitting and receiving antennas at 5.8 GHz (\cite{Pei2021}).

\nomenclature{RIS}{Reconfigurable intelligent surface}
\nomenclature{RFI}{Radio-frequency interference}
\nomenclature{LNA}{Low-noise amplifier}

\chapter{Computational Electromagnetics}
Although the derived results (\cite{Ozgecan2020}) presented in Fig.~\ref{fig:2.2} were not carried out using reliable full-fledged simulation tools (but merely analytical methods), we can attempt to repeat the theoretical derivation and validate the results of the paper, by using a complete 3D simulation utility that will enable us to solve the problem computationally. This approach is commonly followed to design and simulate propagation models, antennas, and other RF networks, as it is the most reliable method of obtaining accurate results.

\section{Introduction to Simulation Techniques}

The techniques used to model and solve complex problems involves the use of computational electromagnetics (CEM). By aiming to solve Maxwell's equations computationally, CEM techniques attempt to provide the radiating characteristics and electromagnetic properties of a modelled structure. Evaluating the elctromagnetic fields throughout the domain of the problem enables the identification and study of the behavior of the electric ($E$-field) and magnetic ($H$-field) components of the electromagnetic waves present in the problem domain. However, because all physical structures consist of an infinite number of points, it is impossible to solve Maxwell's equations across the entirety of a given structure, regardless of its size.

\section{Discretization (Meshing)}

The way around the aforementioned problem is discretization (meshing). Discretization works by splitting the geometry of interest using grids, forming many small cells (meshcells). The greater the number of meshcells, the closer the modelled geometry represents the (actual) physical model. Thus, CEM methods can solve Maxwell's equations at each point in the formed grid (mesh), and yield an approximate (though usually accurate) result of the provided electromagnetic problem.\\

\begin{figure}[htbp]
\centering
\includegraphics[width=\textwidth]{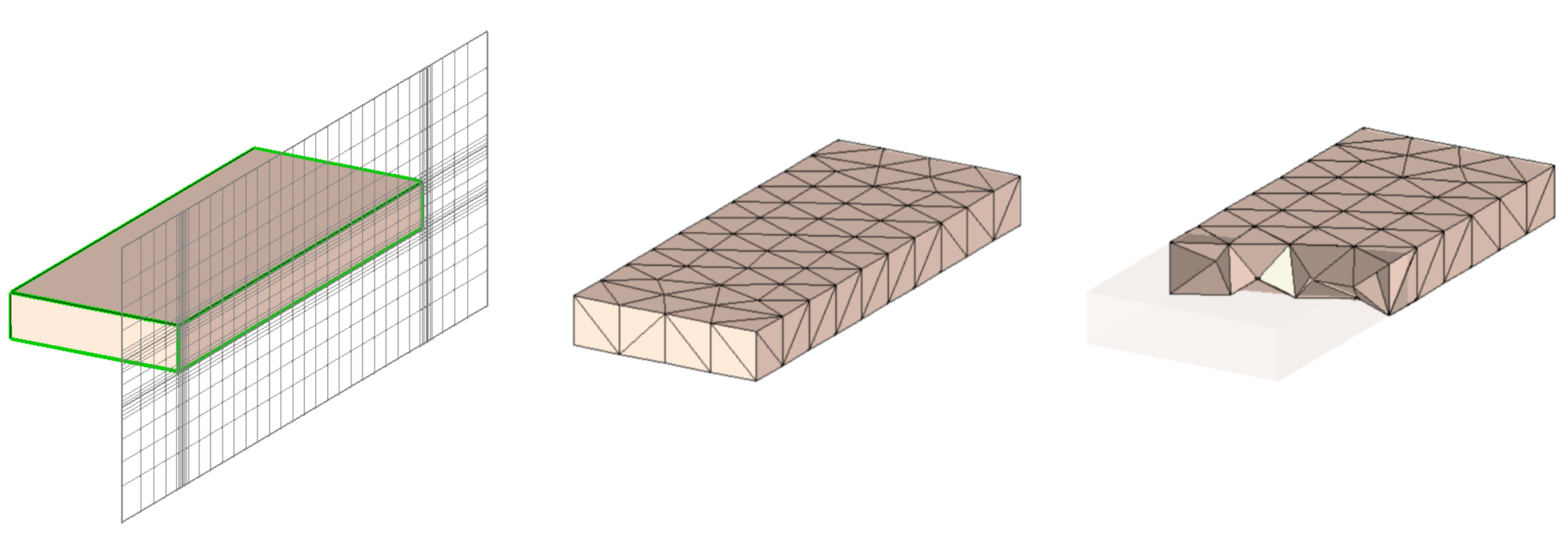}
\caption{Mesh generated and applied to a brick-shaped cuboid. \textit{Left:} generation of a rectilinear mesh consisting of hexahedral cells. \textit{Center:} generation of a tetrahedral mesh, where each meshcell consists of 4 vertices and 6 edges, bounded by 4 triangular faces. \textit{Right:} tetrahedral mesh truncated by a cutting plane, highlighting the irregular connectivity of the inner structure of the unstructured grid.}
\label{fig:mesh}
\end{figure}

Depending on the solver being used, discretization can take various geometrical forms. The two most common types of grids used to solve three-dimensional electromagnetic problems are depicted in Fig.~\ref{fig:mesh}.\\

It is worth noting, that depending on the solver used to tackle a given problem, only a certain type of mesh may be applicable to solve Maxwell's equations on. Furthermore, different each type of mesh offers a variety of advantages and disadvantages.

For example, a rectilinear mesh is well applicable to planar surfaces, such as microstrip designs and wire antennas. However, hexahedral meshcells fail to accurately represent curved structures (including angled\footnote{Straight and right-angled traces and structures are represented by rectilinear grids accurately, but staircase representations are unavoidable when the local geometry of interest extends beyond the axes of the grid planes.} traces and round reflecting structures), and often tend to lead to so-called staircase approximations. While this accuracy-degradation by-product can be minimized with the use of a very fine mesh, this leads to a significantly longer simulation time.\\

On the other hand, the tetrahedral type of mesh deals with curved structures with few restrictions, and can accurately represent complex geometries with a relatively small number of meshcells. Nevertheless, this approach can be very limited when it comes to representing very thin volumes, such as traces and miniature surface mount devices (SMD), as it is geometrically challenging to connect two tetrahedra of disproportionately different sizes.

\section{Solvers}

Although the fundamental theory behind electromagnetic solvers has been well-established for decades, the advances in computational hardware over the last few decades has introduced a variety of new solvers for tackling electromagnetic problems of RF models. The following methods described below are the most common ones used to tackle RF problems.

\subsection{Finite-Difference Time-Domain}

The finite-difference time-domain (FDTD) technique is amongst the most popular methods of solving electromagnetic problems, particularly at radio frequencies. As the name implies, it is a time-domain solver, and simulations can cover a wide frequency range, outputting results for multiple frequencies of interest in every run.\\

Dating over half a century ago, Kane Shee-Gong Yee first introduced the FDTD technique as a method of solving Maxwell's equations (\cite{Yee1966}). Since then, many adaptations and improvements have been proposed and implemented to tackle all sorts of simulation problems. FDTD applications have been broadly demonstrated in several fields, especially around microwave engineering.\\

The FDTD method solves Maxwell's equations in their differential form, and is adaptable to rectilinear grids, consisting of hexahedra meshcells (cuboids).

\subsection{Finite Element Method}

Contrary to the finite-difference time-domain algorithm, the finite element method (FEM) is a frequency-domain solver. While finite element analysis (FEA) is increasingly popular across multiple fields of science, mathematics, and engineering, to tackle a wide variety of prediction problems, it has various advantages and disadvantages in the field of RF and microwave simulations.\\

To begin with, FEM solves differential equations, which, in the field of RF simulations, obviously corresponds to Maxwell's equations. The main considerations simulation engineers take into account when dealing with solver selection for their problem is the type of mesh they would be working with, as well the kind of results they would be viewing.\\

With regard to the discretization process, FEM applies a tetrahedral mesh to the model of interest (as opposed to the FDTD method that uses hexahedral grids). This makes FEA a lot easier to work with complex geometries and curved structures, and provides more accurate results compared to time-domain solvers that struggle with local staircase representations of complex geometries with turns and curves.\\

However, it is worth noting that a major limitation of the finite element method is its output: in order to simulate a wide frequency range and obtain accurate results for multiple frequencies, many individual simulations need to be run. The total number of simulated points, $N_\mathrm{pts}$, for a given frequency range is given by

\begin{align}
    N_\mathrm{pts} = \frac{f_{1}-f_{0}}{\Delta f},
\end{align}

where $f_{1}$ and $f_{0}$ correspond to the highest and lowest excited frequency, respectively, and $\Delta f$ represents the frequency step (expressing the resolution\footnote{Not to be confused with accuracy or precision, the frequency step determines how many points lie between $f_{1}$ and $f_{0}$.} of the simulation).

\subsection{Finite Integration Technique}

Proposed nearly half a century ago (\cite{Weiland1977}), the finite integration technique (FIT) works based on the conservation of energy and charge. As the name hints, this technique makes use of Maxwell's equations in their integral form, applied to the model's mesh. The finite integration technique works with a rectilinear mesh (hexahedral cells), and like the finite-difference time-domain method, FIT can offer results covering a wide frequency range (i.e., provide a continuous output).\\

Due to their intense use in commercial tools and open-source software, FIT algorithms have been improved in several ways; not only in terms of accuracy, but also computational complexity (including more efficient and lighter memory usage). These advantages make FIT a great option for solving electromagnetic problems, even for machines with modest hardware.

\section{Modelling the Optical Response of Surfaces at Sub-lambda Scales}

Given the fundamental simulation methods and techniques are understood, it is now possible to attempt the approach of the reflection problem presented in the study of {\"O}zdogan et al., with the help of computational electromagnetics.

\subsection{Simulation Setup \& Modelling}

To begin with, we can settle on an excitation frequency of 30 GHz. Although this can be parametrized at a later stage, given all dimensions are a function of the wavelength $\lambda$, the precise frequency is of little importance.\\

In order to model the problem, we can first begin by building the planar surface, to which the incident monochromatic electromagnetic wave will be directed. Since the original paper assumes a perfectly square, two-dimensional surface with no thickness, we shall follow suit. The material we assume is a perfect electrical conductor (PEC), with no losses. This ensures the simplicity of the model is as closely representative of the theoretical approach of the paper as possible, while maintaining a minimal simulation time.\\

Due to simulation's nature, a challenging aspect of a problem like this is the incidence angle $\theta_i$, which turns out to be most conveniently aligned with either one of the 6 faces of the boundary box:

\begin{center}
\begin{minipage}{.16\textwidth}
\begin{itemize}
 \setlength\itemsep{0em}
 \item $\mathrm{XY}$
 \item $\mathrm{YZ}$
 \item $\mathrm{XZ}$,
\end{itemize}
\end{minipage}
\end{center}
or, the opposite boundary faces:

\begin{center}
\begin{minipage}{.16\textwidth}
\begin{itemize}
 \setlength\itemsep{0em}
 \item $(\mathrm{XY})'$
 \item $(\mathrm{YZ})'$
 \item $(\mathrm{XZ})'$.
\end{itemize}
\end{minipage}
\end{center}

In other words, because

\begin{align} \label{eq:boundary_face_angle}
    \theta_i \neq \frac{k\pi}{2}, k\in\mathbb{N},
\end{align}

given the plane wave source is most conveniently defined to cover an entire boundary face of our choice, Equation~\ref{eq:boundary_face_angle} implies the incident ray's angle shall be modelled by setting the reflective surface at a slope, equal to $\theta_i$ (relative to the orientation of the plane wave source).\\

\begin{figure}[htbp]
\centering
\includegraphics[width=\textwidth]{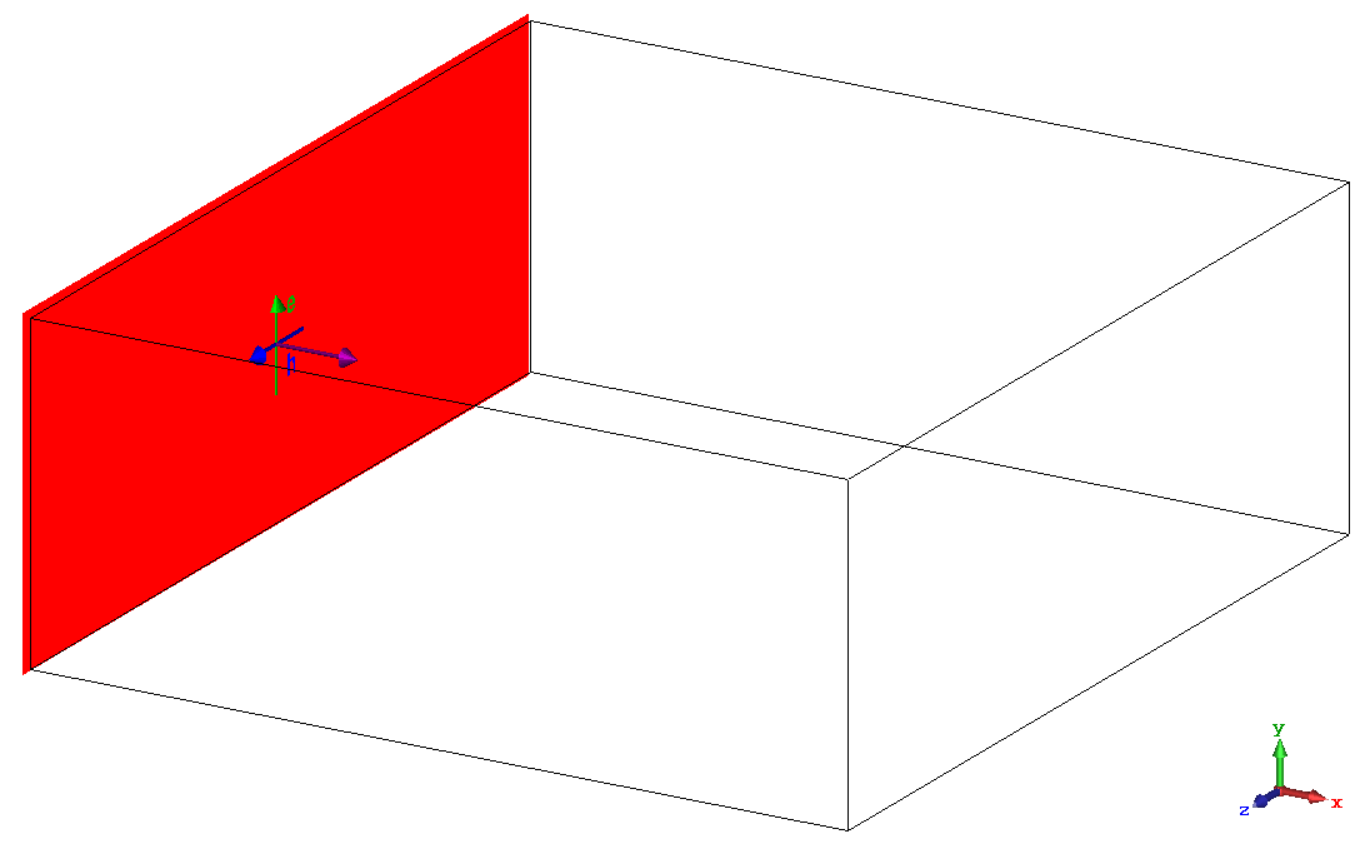}
\caption{Position and orientation of the plane wave source. The electric field vector (denoted by $e$) and magnetic field vector (denoted by $h$) highlight the vertical polarization of the electromagnetic waves to be excited upon the initiation of the simulation.}
\label{fig:planewave}
\end{figure}

As shown in Fig.~\ref{fig:planewave}, the plane wave source has been defined at the $\mathrm{YZ}$ face, and the polarization has been set to linear (vertically polarized).

The purple vector in Fig.~\ref{fig:planewave} indicates the Poynting vector, which provides the direction of propagation (\cite{Poynting1884}) for a transverse electromagnetic wave, given by the cross product

\begin{align} \label{eq:poynting}
\mathbf{S} =\mathbf{E} \times \mathbf{H}.
\end{align}

\subsubsection{Boundaries}

Assuming a frequency of 30 GHz, the wavelength we are working with is $\sim10\mathrm{\ mm}$. By default, the boundary conditions are set to open on all six sides of the boundary box, with an additional fraction of a wavelength added to provide more space around the edges of the structure, e.g.

\begin{align} \label{eq:quarter_wavelength}
\frac{1}{4} \lambda.
\end{align}

This helps ensure electromagnetic waves have enough room to properly form before being ``absorbed'' by the sides of the boundary box. The goal of a sufficiently large added space is the assurance of accurate results, representative of the true behavior of the electromagnetic radiation present in the model, and its interactions with physical structures.\\

\begin{figure}[htbp]
\centering
\includegraphics[width=\textwidth]{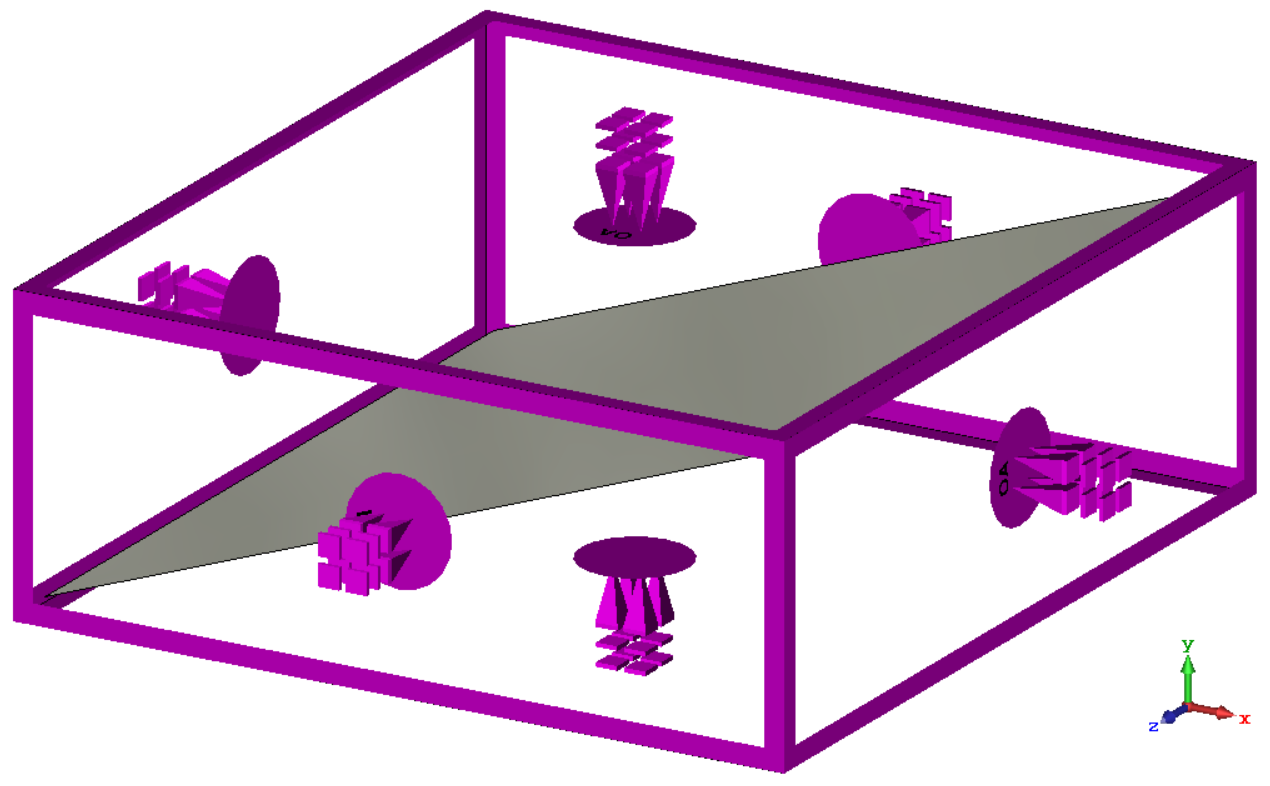}
\includegraphics[width=0.9\textwidth]{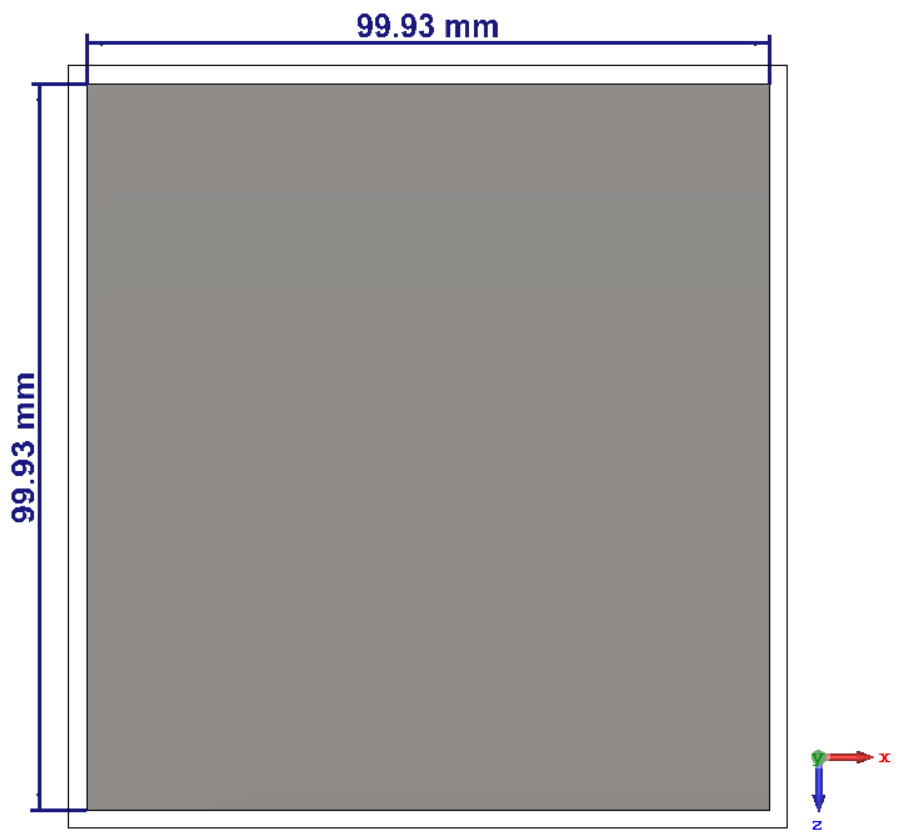}
\caption{\textit{Top:} boundary conditions of the simulation model. The purple symbols indicate absorbers of electromagnetic radiation, present on all six sides of the boundary box. \textit{Bottom:} top view of the 2D atom. Notice the additional quarter wavelength space extended from the edges of the atom's surface.}
\label{fig:open_add_space}
\end{figure}

Fig.~\ref{fig:open_add_space} shows the open conditions of the boundary box, along with the top view of the atom's 2D planar structure. For the initial simulation, a width and length of $10 \lambda$ were set.

\subsubsection{Parametrization}

In electromagnetic simulations of RF systems, it is often useful to parametrize values related to the geometry of the structure, as well as excitation properties. This makes the model easily adjustable, enabling a modular and versatile simulation environment, without the need of drastically redefining the dimensions of geometrical structures and simulation-related settings. Table~\ref{tab:parameters} lists the parameters set for the simulation model, along with a description for each variable.

\begin{table}[h!]
 \caption{Defined model parameters.}
\label{tab:parameters}
\centering
\begin{tabular}{|c | c | c | c|}
 \hline
 \textbf{Name} & \textbf{Expression} & \textbf{Value} & \textbf{Description} \\ [0.2ex] 
 \hline\hline
 \texttt{c} & \texttt{299792458} & $299792458$ & Speed of light in vacuum, $c$ [$\mathrm{m}/\mathrm{s}$] \\
 \hline
 \texttt{frequency} & \texttt{30} & $30$ & Excitation frequency, $f$ [GHz] \\
 \hline
 \texttt{lambda} & \texttt{c*1e3/frequency*1e-9} & $\sim9.99$ & Wavelength, $\lambda$ [mm] \\
 \hline
 \texttt{scale} & \texttt{10} & $10$ & Parametric scaling factor, $s$ \\
 \hline
 \texttt{width} & \texttt{scale*wavelength} & $\sim99.9$ & Surface width, $w$ [mm] \\ 
 \hline
 \texttt{length} & \texttt{scale*wavelength} & $\sim99.9$ & Surface length, $l$ [mm] \\
 \hline
 \texttt{theta} & \texttt{60} & $60$ & Incident plane wave angle [deg] \\
 \hline
\end{tabular}
\end{table}

\subsubsection{Discretization}

\begin{figure}[htbp]
\centering
\includegraphics[width=0.77\textwidth]{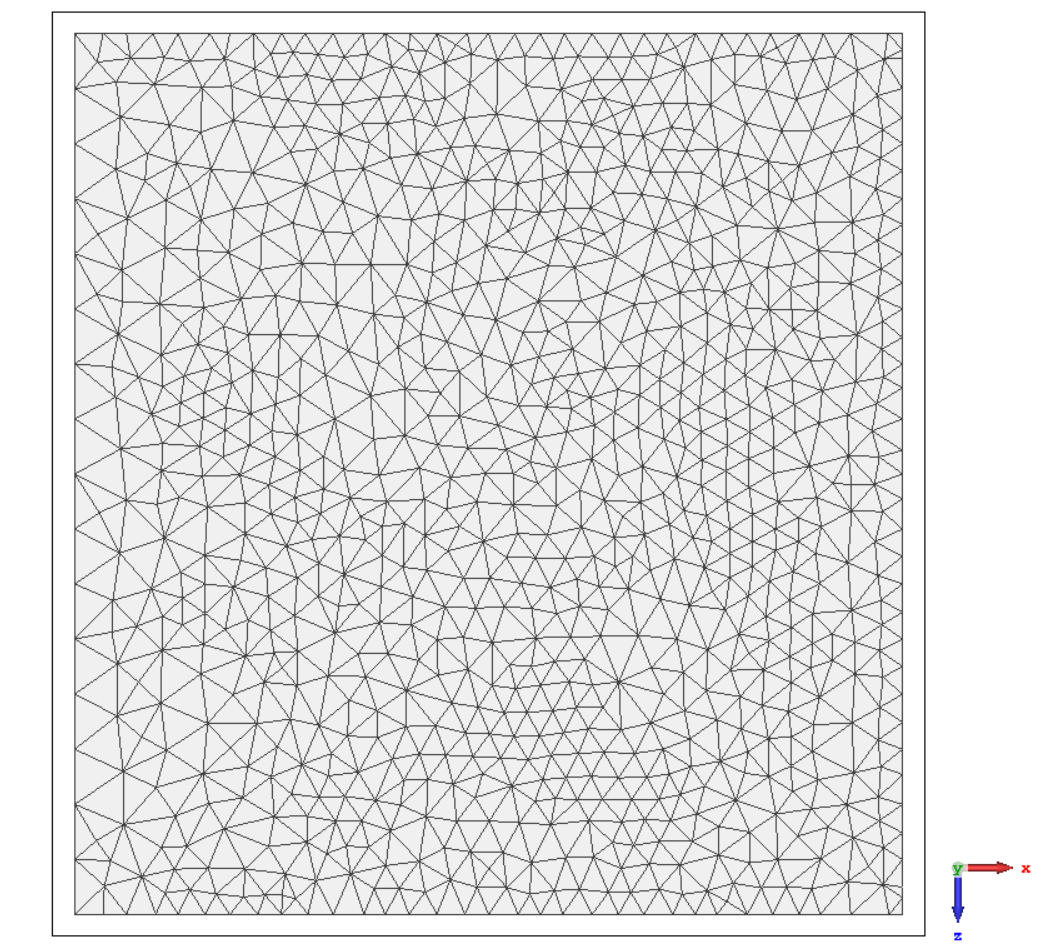}
\caption{Discretization of a two-dimensional atom into triangle-shaped meshcells.}
\label{fig:fig_mesh_atom}
\end{figure}

Before proceeding to the simulation, we shall first ensure the mesh of our model is fine enough, but not too coarse. Since we are dealing with a 2D surface (and not a conventional 3D structure), the meshcells are now two-dimensional triangles, instead of three-dimensional tetrahedra. The mesh structure is shown in Fig.~\ref{fig:fig_mesh_atom}.

\begin{figure}[htbp]
\centering
\includegraphics[width=\textwidth]{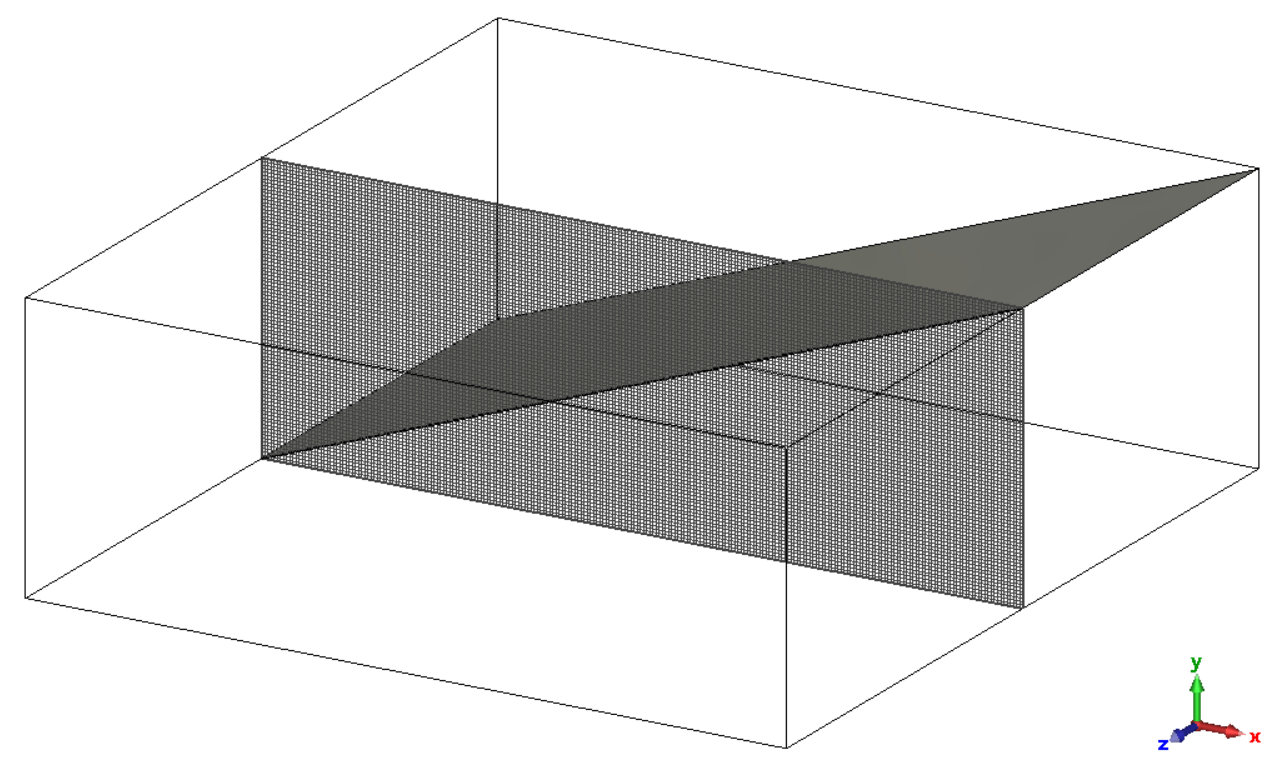}
\caption{Hexahedral mesh employed by the (time-domain) finite integration technique solver. Note the staircase representation provided by the non-parallel slope of the atom's surface. Additionally, an unnecessary amount of unintentional (but still obligatory) detail is dedicated beyond the atom (free space), yielding a massive amount of meshcells. As a result, the soaring computational time introduced by this type of solver for this particular electromagnetic problem encourages us to go with the (frequency-domain) finite element method instead.}
\label{fig:fig_hex_mesh_atom}
\end{figure}

Notice that, if we were to go with a hexahedral mesh (like in Fig.~\ref{fig:fig_hex_mesh_atom}), we would deal with the staircase problem discussed earlier, which would yield results of lower accuracy.\\

To ensure the highest possible accuracy in the simulation, we employ a 3rd-order finite element method solver to yield results that are interpretable in a reliable manner.

\subsection{Simulation Results}

\begin{figure}[htbp]
\centering
\includegraphics[width=\textwidth]{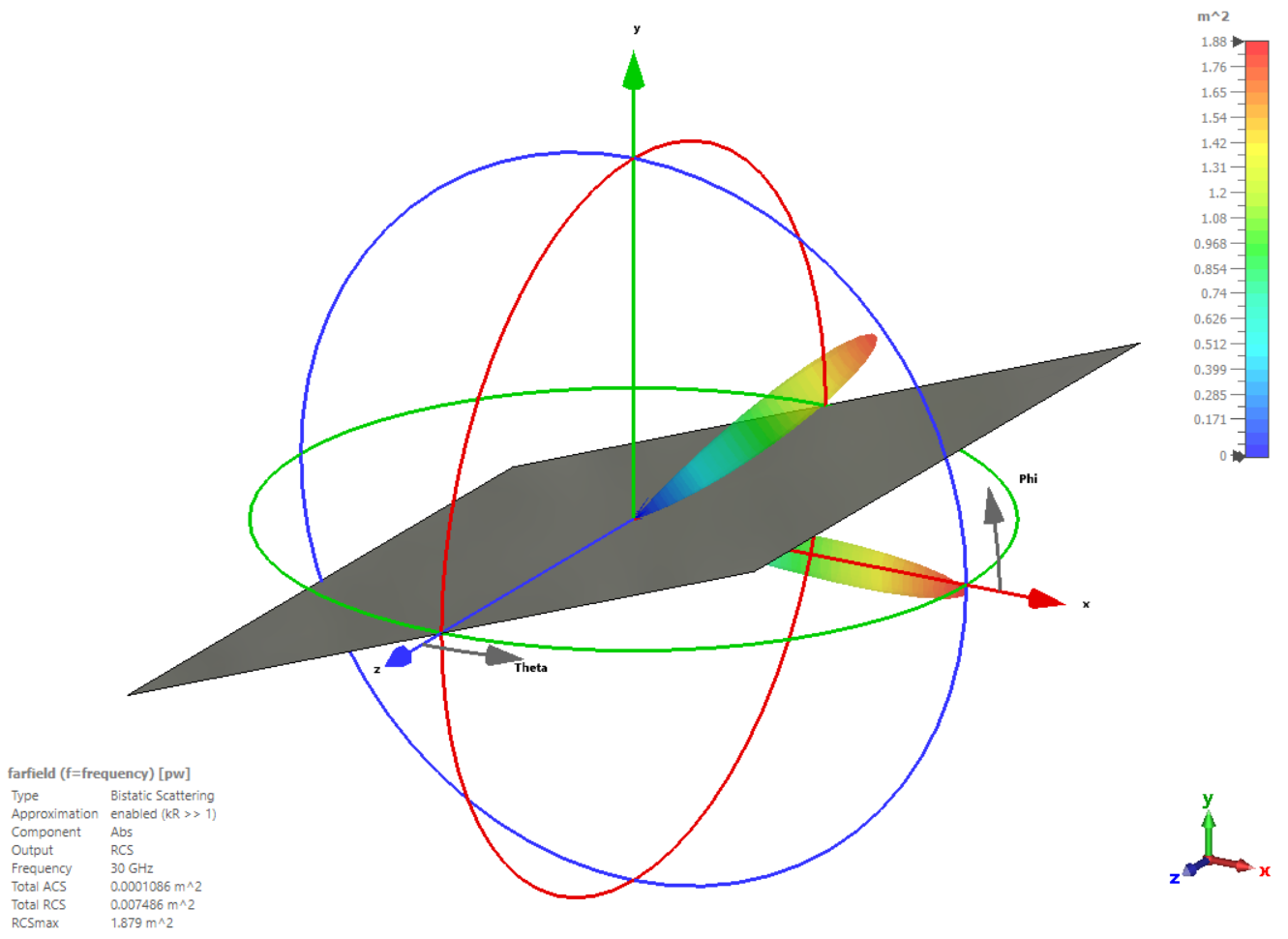}
\caption{Results of the first simulation. The atom has a width and length equal to 10 wavelengths, at the excitation frequency of 30 GHz. This corresponds to $a=b=10$ cm. The scaling of the radar cross-section is displayed in linear units (m$^2$ instead of dBms). This sacrifice in dynamic range helps in the visualization by highlighting the direction of the reflected (scattered) lobe with better precision.}
\label{fig:fig_diffraction}
\end{figure}

The preliminary results of the first computation, with the simulation set up as described in the aforementioned paragraphs are presented in Fig.~\ref{fig:fig_diffraction}. The intensity is given by the so-called radar cross-section (RCS), in units of square meters ($\mathrm{m}^2$ or dBsm). The three-dimensional RCS, $\sigma$, is given by the limit

\begin{align} \label{eq:rcs_power_density}
\sigma =\lim _{r\to \infty }4\pi r^{2}{\frac {S_{s}}{S_{i}}},
\end{align}

where $S_{i}$ is the incident power density measured at the atom, and $S_{s}$ is the power density scattered, observed at distance $r$ from the atom (\cite{Balanis2012}). We may also express this in terms of the (far-field) intensity of the electric fields (\cite{Knott2004}), i.e.,

\begin{align} \label{eq:rcs_electric_field}
\sigma =\lim _{r\to \infty }4\pi r^{2}{\frac {|E_{s}|^{2}}{|E_{i}|^{2}}},
\end{align}

where $E_{s}$ is the intensity of the scattered electric field and $E_{i}$ is the intensity of the incident electric field.

\begin{figure}[htbp]
\centering
\includegraphics[width=\textwidth]{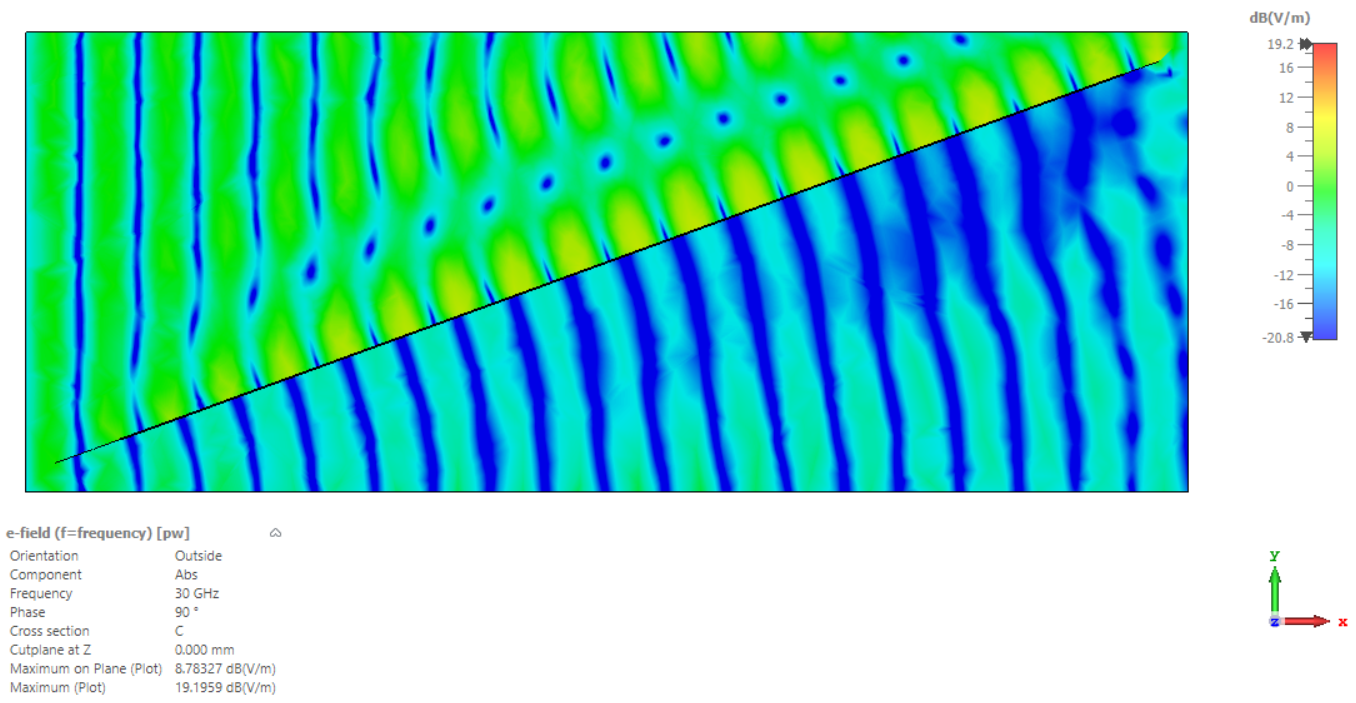}
\caption{Cross-section (side view) of the E-field distribution on the atom at an arbitrary phase of $90^{\circ}$, displayed in units of $\mathrm{dB}(\mathrm{V}/\mathrm{m})$.}
\label{fig:fig_efield_diffraction}
\end{figure}

Notice that, while the plane wave is scattered off of the atom's surface at a reflection angle equal to the incident wave's angle, following Snell's law (Equation \ref{eq:2.3}) as expected, an additional lobe appears to be formed underneath the atom's surface. If we investigate the E-field distribution (Fig.~\ref{fig:fig_efield_diffraction}) of the simulation, we can directly see and understand the precise reason this happens.\\

As mentioned earlier in Equation~\ref{eq:quarter_wavelength}, there is an added space between the edge of the modelled structure and the plane wave source. This unnecessary space introduces sufficient room for diffraction effects to form, as the gap between the edge and the side of the boundary box is exposed to the incident electromagnetic radiation. Hence, the diffracted waves (which are visibly ``climbing up'' the underside of the atom's surface) are undoubtedly the source of the second lobe seen in Fig.~\ref{fig:fig_efield_diffraction}.\\

\begin{figure}[htbp]
\centering
\includegraphics[width=\textwidth]{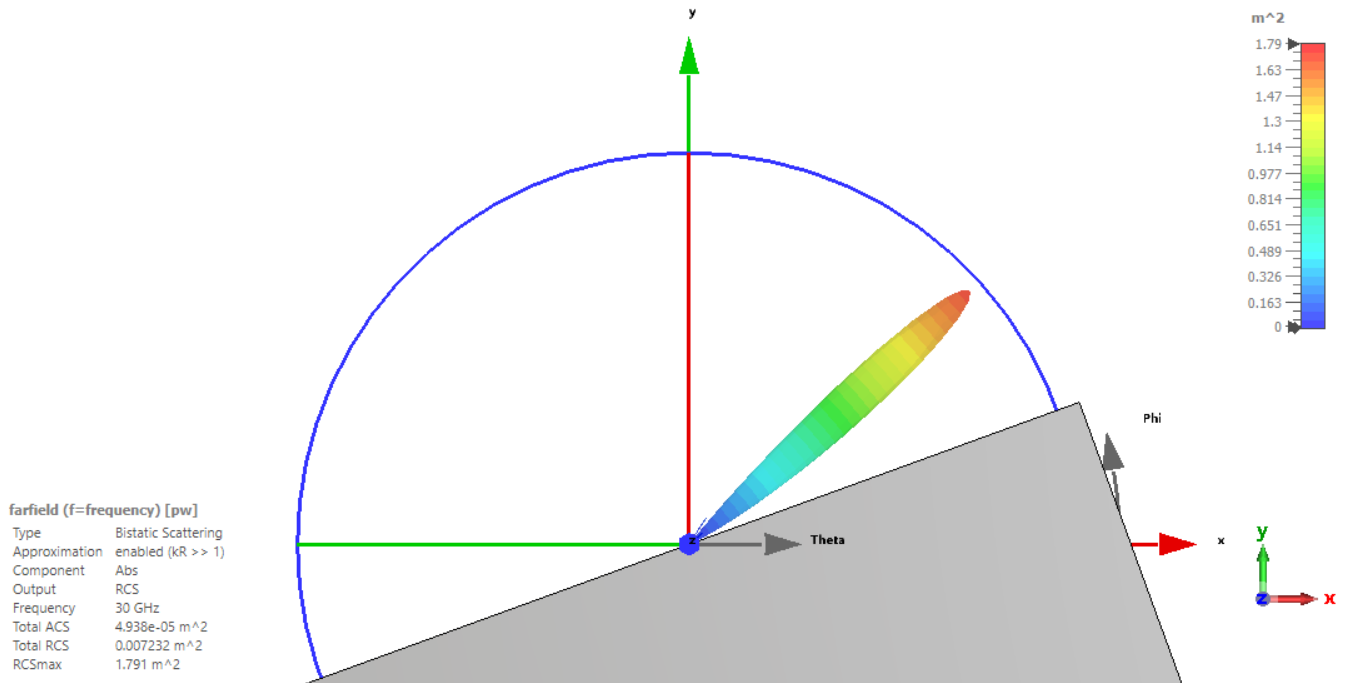}
\caption{RCS result of the second simulation. The atom has a width and length equal to 10 wavelengths, at the excitation frequency of 30 GHz. Note that the second lobe is now completely eliminated.}
\label{fig:fig_10lambda}
\end{figure}

\begin{figure}[htbp]
\centering
\includegraphics[width=\textwidth]{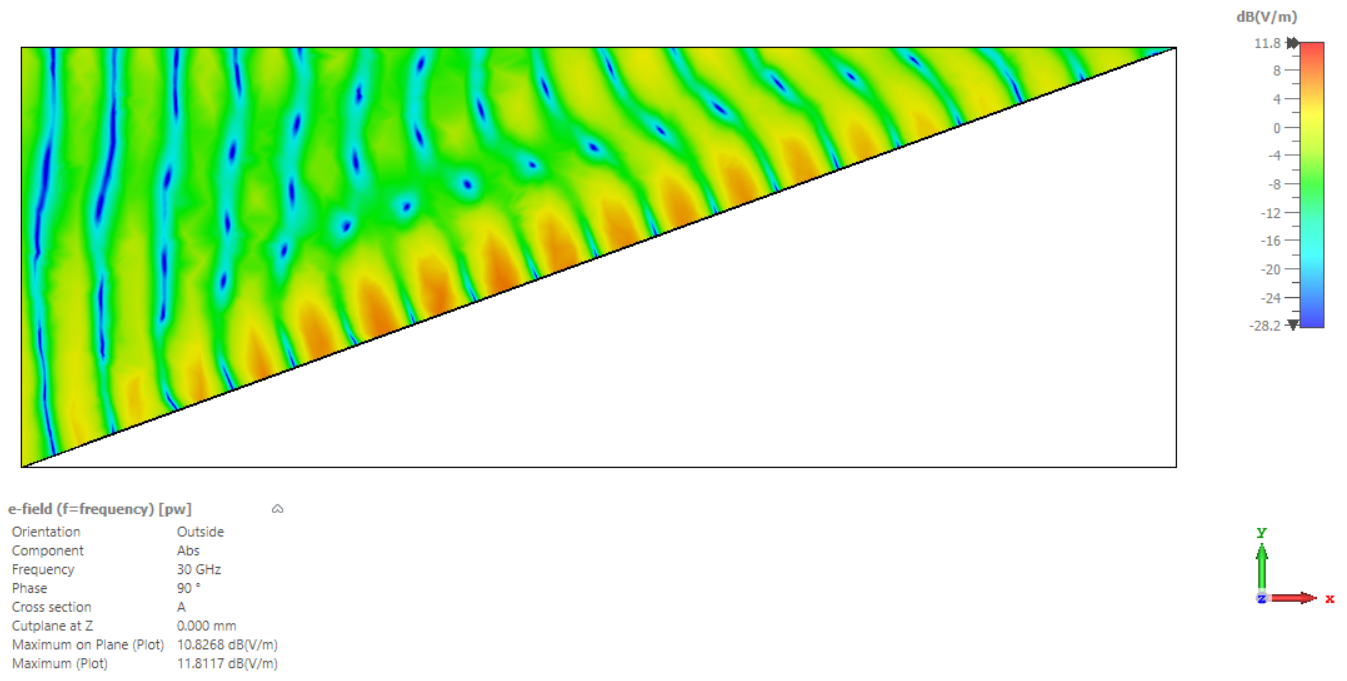}
\caption{Cross-section (side view) of the E-field distribution on the atom at an arbitrary phase of $90^{\circ}$, displayed in units of $\mathrm{dB}(\mathrm{V}/\mathrm{m})$. The electromagnetic effect of the extrusion of the surface into a three-dimensional volume is visible by the presence of the E-field intensity, solely on the upper side of the atom's surface.}
\label{fig:fig_10lambda_efield}
\end{figure}

In order to resolve this issue and only preserve the main lobe of interest (the one expected from Snell's law), we can ensure that the value of Equation~\ref{eq:quarter_wavelength} is nullified. In other words, while the boundary conditions shall remain open, the space between the edge of the surface and the plane wave should be neglected. After all, the wavefronts of the radiation source need no space to ``mature'' into a different shape or form (i.e., since we are dealing with plane waves, the concept of the far-field region does not apply). To further ensure no energy is coupled to the underside of the atom, we can extrude the two-dimensional surface and transform it into a three-dimensional volume. Fig.~\ref{fig:fig_10lambda} shows the RCS of the second simulation, with the ``beam-split'' issue resolved. while Fig.~\ref{fig:fig_10lambda_efield} demonstrates the E-field distribution after the elimination of the diffraction effect.\\

Despite the boundary and structural modifications, however, it was found that the secondary lobe was still present. While the presumed origin of this lobe is the plane wave source, we can completely ignore it and solely focus on the primary lobe of interest. This enables us to study the shape of the reflection pattern, which has been the initial goal we have been interested in tackling.

\subsection{Evaluating the Numerical Simulation Against the\\Reference Analytical Method}

To verify the results presented in previous work (\cite{Ozgecan2020}), we have tackled the presented electromagnetic problem using a numerical technique. The simulation was accomplished using the accurate 3rd-order finite element method technique. This allows us to confirm the analytical method the authors approached is indeed valid. Due to the limitations in computing resources, however, simulating an atom surface with dimensions beyond a certain point $(a=b>10\lambda)$ was impractical, because of the exceedingly large mesh of the model.\\

Thus, to present a valid comparison between this numerical simulation and the study used as a reference for the analytical method, the same dimensions shall be used. The paper provides results for the following set of dimensions:

\begin{align} \label{eq:paper_dimensions}
a=b=\{\lambda/2, 10\lambda, 50\lambda\}.
\end{align}

By transforming the aforementioned set into a collection of dimensions we can simulate with the limited computing resources under consideration, we get the following set:

\begin{align} \label{eq:our_dimensions}
a=b=\{\lambda/2, 5\lambda, 10\lambda\}.
\end{align}

We begin by reproducing the analytical solution described in {\"O}zdogan et al., based on the provided open-source code\footnote{\url{https://github.com/emilbjornson/IRS-modeling/blob/master/plotFigure5.m}} used to generate the original figures. By altering lines 33 and 34, we can transform Set (\ref{eq:paper_dimensions}) to the desired Set (\ref{eq:our_dimensions}). The code segment ends up like so:\\

\begin{lstlisting}[language=Python]
% #Sizes of the surfaces for comparison
a=[0.5,10,50]*lambda; 
b=[0.5,10,50]*lambda;
\end{lstlisting}

\begin{figure}[htbp]
\centering
\includegraphics[width=\textwidth]{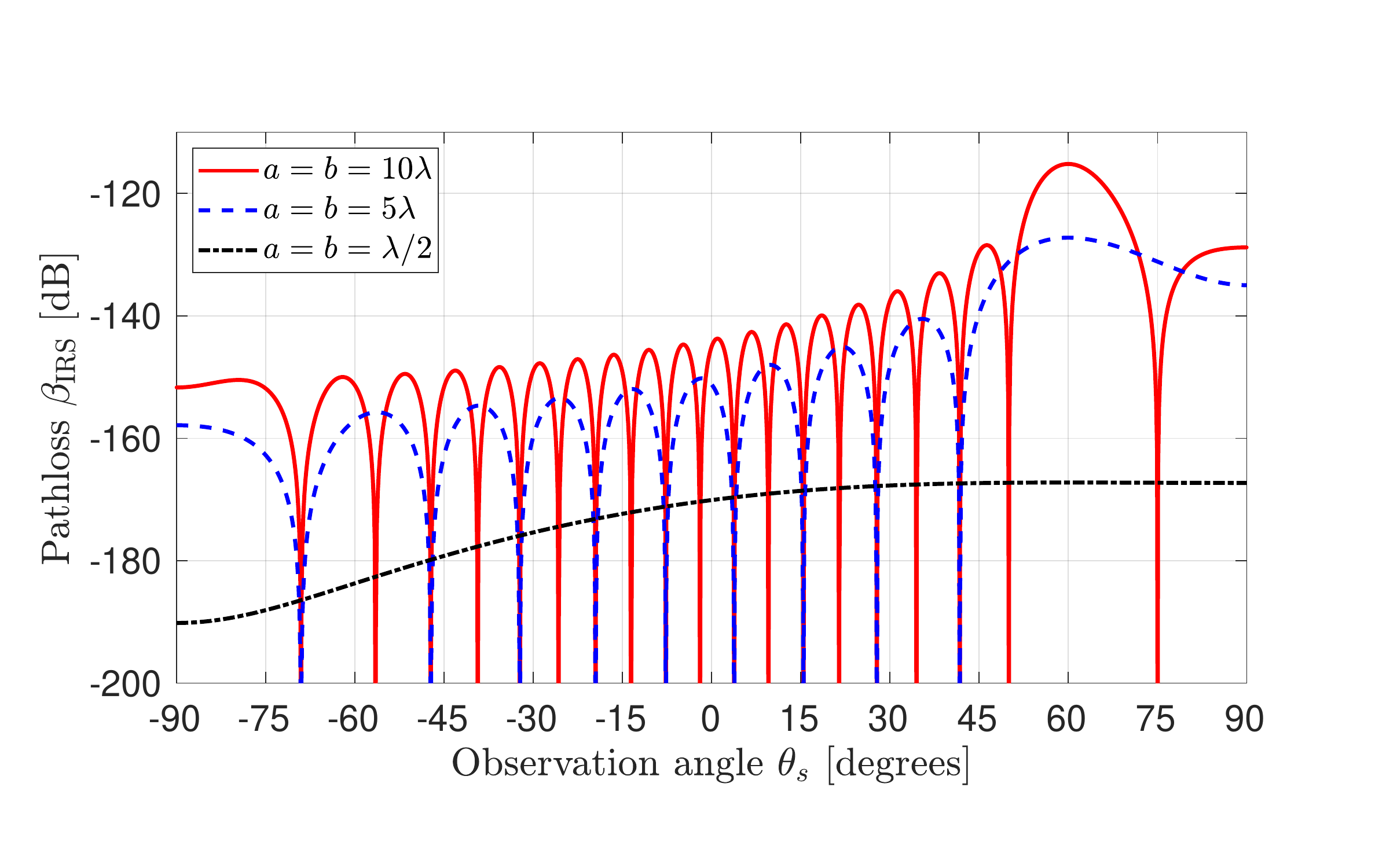}
\caption{Analytical solution for 30 GHz, reshaped from the original study. The plot corresponds to three discrete dimension samples, as defined in Set (\ref{eq:our_dimensions}). The angle of incidence is $30^{\circ}$, and, according to Snell's law, the expected angle of reflection is $60^{\circ}$.}
\label{fig:fig_matlab}
\end{figure}

\mbox{}\\The output is shown in Fig.~\ref{fig:fig_matlab}. The initial observations derived from the figure are the following:
\begin{enumerate}
   \item The peak amplitude (global maximum) is consistently at $60^{\circ}$, regardless of $a=b$. The result is thus consistent with Snell's law, irrespective of physical dimensions;
   \item The larger the size of the reflecting surface of the atom (relative to the wavelength) becomes, the greater the peak amplitude of the reflected wave gets;
   \item No grating lobe is formed, and the amplitude of each sidelobe increases as $\theta_s \to \theta_r$;
   \item The half-power beamwidths of the main lobe and sidelobes narrow as $p \to \infty$; and
   \item The number of sidelobes, $N_\mathrm{sidelobes}$, grows with the wavelength's coefficient, proportionally:
 \end{enumerate}

\begin{align} \label{eq:proportionality}
N_\mathrm{sidelobes}=2p-1,
\end{align}

where $p \in \mathbb{R}_{>0}$ is the coefficient of the wavelength. The introduction of $-1$ accounts for the main lobe, which should not be considered a sidelobe.\\

Of course, points 2, 4, and 5 are expected results derived from the law of conservation of energy.

\begin{figure}[htbp]
\centering
\includegraphics[width=\textwidth]{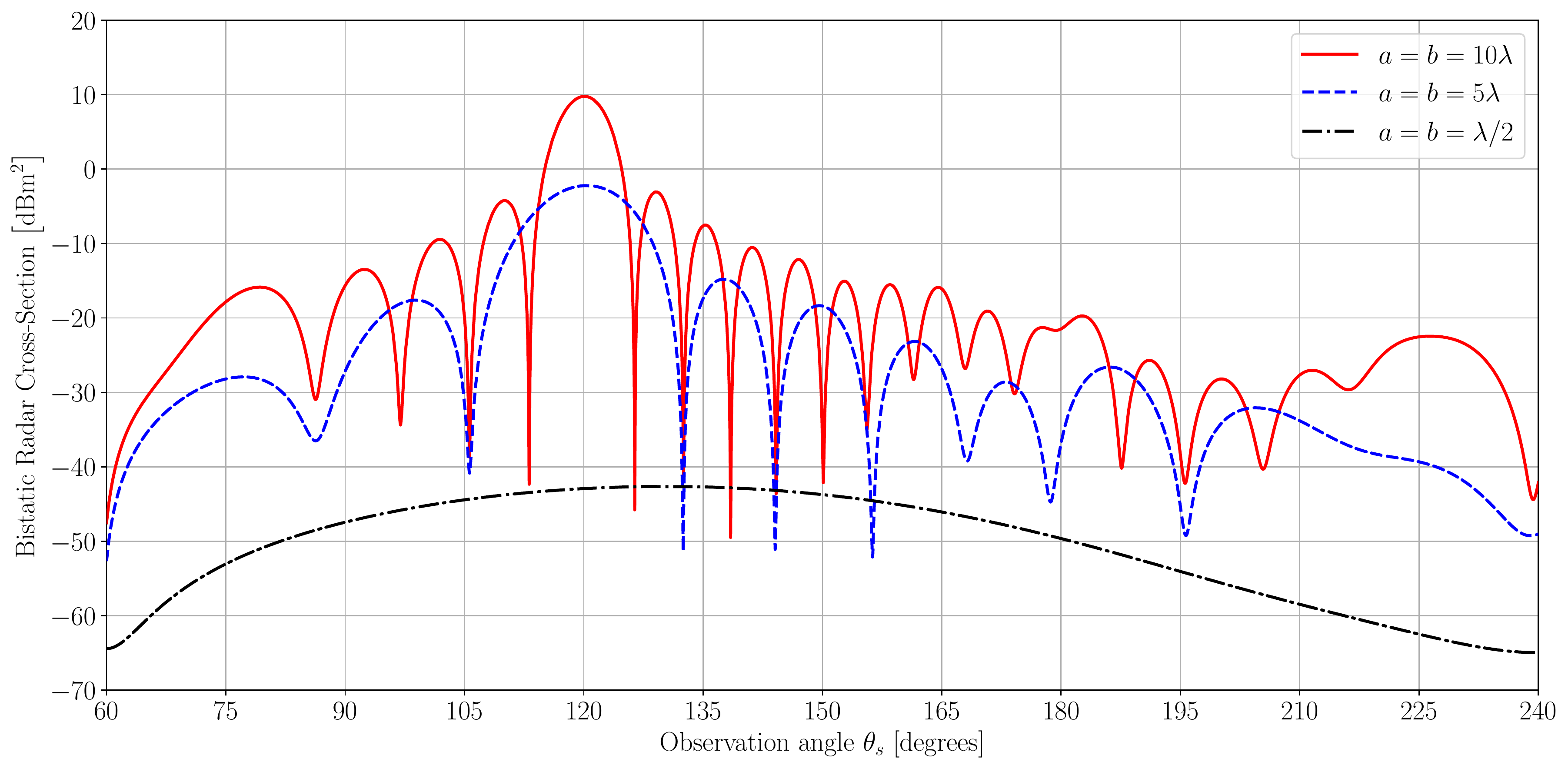}
\caption{Numerical solution for 30 GHz, based on the finite element method. The set again corresponds to Equation~\ref{eq:our_dimensions}, matching the dimensions presented in the previous figure. Instead of pathloss, the radar cross-section is displayed.}
\label{fig:fig_fem_mpl}
\end{figure}

If we plot the results from the numerical simulation, conducted with the help of the finite element method, we get the results shown in Fig.~\ref{fig:fig_fem_mpl}.\\

It can be seen that our simulation is in general agreement with the aforementioned paper, highlighting the validity of the study's analytical approach. Furthermore, it can be seen that several characteristics match between the expected result and the simulation. Firstly, the peak value (global maximum of RCS) aligns with Snell's law. Secondly, Equation~\ref{eq:proportionality} is in agreement with the number of sidelobes present in all three curves. Thirdly, the difference (or ratio, if expressed in linear scale) is consistent between the analytical and the numerical method.  Last but not least, the half-power beamwidth of the main lobe and sidelobes narrow with the growth of the $p$ coefficient.\\

Thus, the key characteristics highlighted in the observation of the study's analytical results are consistent with our simulations. We therefore conclude that the analytical derivation provided in the paper is highly accurate, and vastly less computationally expensive compared to numerical approaches involving the finite element method, finite integration technique, finite-difference time-domain, or other electromagnetic solvers.

\subsection{Implications for the Operation of RIS}

In the previous subsection, we demonstrated the optical response of surfaces, across a wide range of scales: $\lambda/2$, $5\lambda$, and $10\lambda$. What Fig.~\ref{fig:fig_matlab} and \ref{fig:fig_fem_mpl} show is that the smaller the size of the surface is (in terms of the operating wavelength), the wider the resulting lobe gets. This is a very critical observation, as it highlights the primary reason for why reconfigurable intelligent surfaces tend to work at such small element scales. If the elements were responding to incident plane waves in a specular manner, then reconfigurable intelligent surfaces could not behave in the desired way, i.e., reflect the incident wave at a particular direction (angle) of interest.\\

Altogether, however, the total re-radiated power from a large number of atom elements is ultimately such, that based on the impedance of each atom (and thus the phase response each element exhibits), the direction of the scattered wave is reconfigurable, and the total intensity requirement is met through constructive interference; similar to the operation of a conventional phased array.

\nomenclature{CEM}{Computational electromagnetics}
\nomenclature{dBsm}{Decibel square meters}
\nomenclature{SMD}{Surface-mount device}
\nomenclature{FDTD}{Finite-difference time-domain}
\nomenclature{FEM}{Finite element method}
\nomenclature{FEA}{Finite element analysis}
\nomenclature{FIT}{Finite integration technique}
\nomenclature{PEC}{Perfect electric
\nomenclature{RCS}{Radar cross-section}
conductor}

\chapter{Efficient Non-Uniform Structured Mesh Generation Algorithm for Computational Electromagnetics}
This chapter consists of a preprint of the Communication article entitled `Efficient Non-Uniform Structured Mesh Generation Algorithm for Computational Electromagnetics,' which deals with the problem of geometry discretization in the field of numerical electromagnetics.

\section{Abstract}

Despite the rapidly evolving field of computational electromagnetics, few open-source tools have managed to tackle the problem of automatic mesh generation for properly discretizing the problem of interest into a finite set of elements (cells). While several mesh generation algorithms have been established in the field of computational physics, the vast majority of such tools are targeted solely towards tetrahedral mesh formation, with the intended primary application being the finite element method. In this work, a computationally efficient non-uniform structured (rectilinear) mesh generation algorithm for electromagnetic simulations is presented. We examine the speed, performance and adaptability against previous work, and we evaluate its robustness against a complex geometry case with a commercially-generated grid. The mesh and simulation results produced using the generated grids of the proposed method are found to be in solid agreement.

\section{Introduction}

{T}{he} field of computational electromagnetics is a rapidly developing area, with many applications revolving around the design of electronics, RF networks, antennas, propagation models, and more.\\ 

The role of computational electromagnetics is to tackle electromagnetic problems that are too complex to be solved using analytical solutions. To achieve this, the geometry of a model is fed to a solver engine, which attempts to solve Maxwell's equations across the entire model, observing the electromagnetic response of the system (\cite{davidson2010}). However, because geometries of electromagnetic devices can take all sorts of complex shapes and forms, and more importantly, because it is simply impossible to solve Maxwell's equations on an infinite number of geometry points, the input geometry must undergo discretization. This way, the number of operations becomes finite, and the problem becomes computationally feasible to tackle.\\

While there are various numerical methods to tackle such electromagnetic problems, the most popular techniques include the finite-difference time-domain (FDTD) (\cite{yee1997}), the method of moments (MoM) (\cite{mom}), and the finite element method (FEM) (\cite{fem}). All of these techniques come with a variety of advantages and disadvantages (\cite{solver_comparison}), but time-domain methods like FDTD are established to be particularly robust when dealing with problems in the field of microwave engineering. This is because unlike frequency-domain methods like MoM and FEM, FDTD's theory of operation is based on stimulating a broadband excitation signal, and observing the system's response across the entire frequency range of interest. This is partially achieved using Fourier transforms, converting time-domain signals to frequency-domain spectra.\\

However, despite FDTD's increasing popularity over the last few decades, few studies have attempted to tackle the problem of geometry discretization using open-source tools. The most notable of which is AEG Mesher (\cite{aegmesher}), which proposes a method of generating three-dimensional rectilinear grids from unstructured tetrahedral meshes, produced using tools like \texttt{Gmsh} (\cite{gmsh}). Despite the novelty of AEG Mesher's technique however, we identify a set of drawbacks that unfortunately introduce certain limitations we aim to address in this paper.\\

Namely, the grid generation algorithm is fairly slow. This is because the formation of an unstructured tetrahedral mesh (which is a computationally expensive task) is a prerequisite. Additionally, the employed algorithm appears to include processes involving lots of computations, and the rectilinear mesh generation code itself could potentially benefit from better CPU utilization using multiprocessing.\\

Furthermore, a major restriction AEG Mesher imposes on geometry discretization is the lack of non-uniform grid generation. While a proof of concept has been demonstrated on the original study, it is yet to be fully integrated into the package for standard applications. This is arguably among the most critical limitations of the aforementioned package, as restricting a model to uniform cuboid cells can yield an unnecessarily massive grid for the FDTD engine to deal with, ultimately leading to suboptimal solver performance.
\pagebreak

In this work, we present an open-source non-uniform rectilinear grid generation tool, that is easily applicable to solvers like FDTD, and is up to hundreds of times faster compared to previous work.\footnote{A fighter jet model (\url{https://github.com/flintoftid/aegmesher/blob/master/examples/jet/Jet.stl}) was used as a reference for the benchmark, yielding a meshing time of $<$0.7 s compared to AEG Mesher's 95 s. Due to the unstructured mesh-generation prerequisite, this speedup ratio is expected to further scale when the geometry is input as a set of 3D solids instead of a plain and simple tetrahedral mesh.} In Section II, we lay out the the data structures and computational considerations of the proposed algorithm, and proceed to describe how the parameters of the code affect the produced grid (Section III). Section IV highlights the implementation of a tree-traversal algorithm to assist in the CAD preprocessing segment of the process. The results of an example case are presented in Section V, where the accuracy is evaluated against commercially-generated grids. After concluding and summarizing our work (Section VI), further enhancements involving mesh classification techniques are proposed as future work (Section VII).\\

\section{Computational Considerations}

In order for the proposed meshing technique to be computationally efficient, the employed algorithm shall make use of appropriate data structures and computational operations, such that the time and space complexity is minimized. This is particularly important for large and complex models consisting of numerous vertices, demanding more operations to process and refine accordingly.

Due to its simple syntax and modularity, the programming language of choice for the implementation of the meshing algorithm is Python. Not only does Python significantly accelerate the development process---enabling quick tests and performance evaluations---but it also provides a seamless integration with the \texttt{CadQuery} Python package (\cite{cadquery}). This library simplifies CAD preprocessing, supporting the input requirements of the core segment of the meshing algorithm.

\subsection{Data Structures}

Despite its simple syntax, Python's simplicity comes with a major drawback. When it comes to lists, each item's type (\textit{boolean}, \textit{integer}, \textit{float}, \textit{string}, etc.) can be arbitrary, making lists a heterogeneous data structure. While this introduces great flexibility, it may also greatly degrade the computational efficiency of a function, as each list element constructs a separate Python object. This differs from low-level languages like C, where the elements are restricted to the predefined type of the array.\\

In order to tackle this issue, we employ the NumPy package (\cite{numpy}), which is a library designed for high-performance scientific computing, and is particularly robust with the introduction of the \texttt{ndarray} object. Unlike Python lists, the size of NumPy arrays is static, consisting of items of the same data type. Furthermore, many operations (provided as a wrapper for low-level code) are carried out by pre-compiled code, imitating the speedy functionality of low-level languages.

\subsection{Further Speedup Optimization Attempts}

Further attempts were carried out to attempt to minimize the array operations, such that mesh generation of even complex geometries could be concluded in milliseconds rather than seconds. Various techniques were tested, including the efficient use of threads, \texttt{Cython} (designed to translate Python source code into optimized and compiled C/C\texttt{++} code) (\cite{cython}), as well as \texttt{Numba}: a just-in-time compiler for accelerating Python and NumPy by converting the source to machine code (\cite{numba}).\\

However, neither of the aforementioned attempts were found to be sufficiently easy to adapt the original code to, and were thus not implemented. In 2/3 cases, implementations were successfully set up to run, but no meaningful improvement was observed. Considering the vast majority of the computing time is spent by the solver, efforts to bring the meshing time further down were discontinued, as the initial algorithmic approach was already fast.

\begin{figure}
  \begin{center}
  \includegraphics[width=0.48\textwidth]{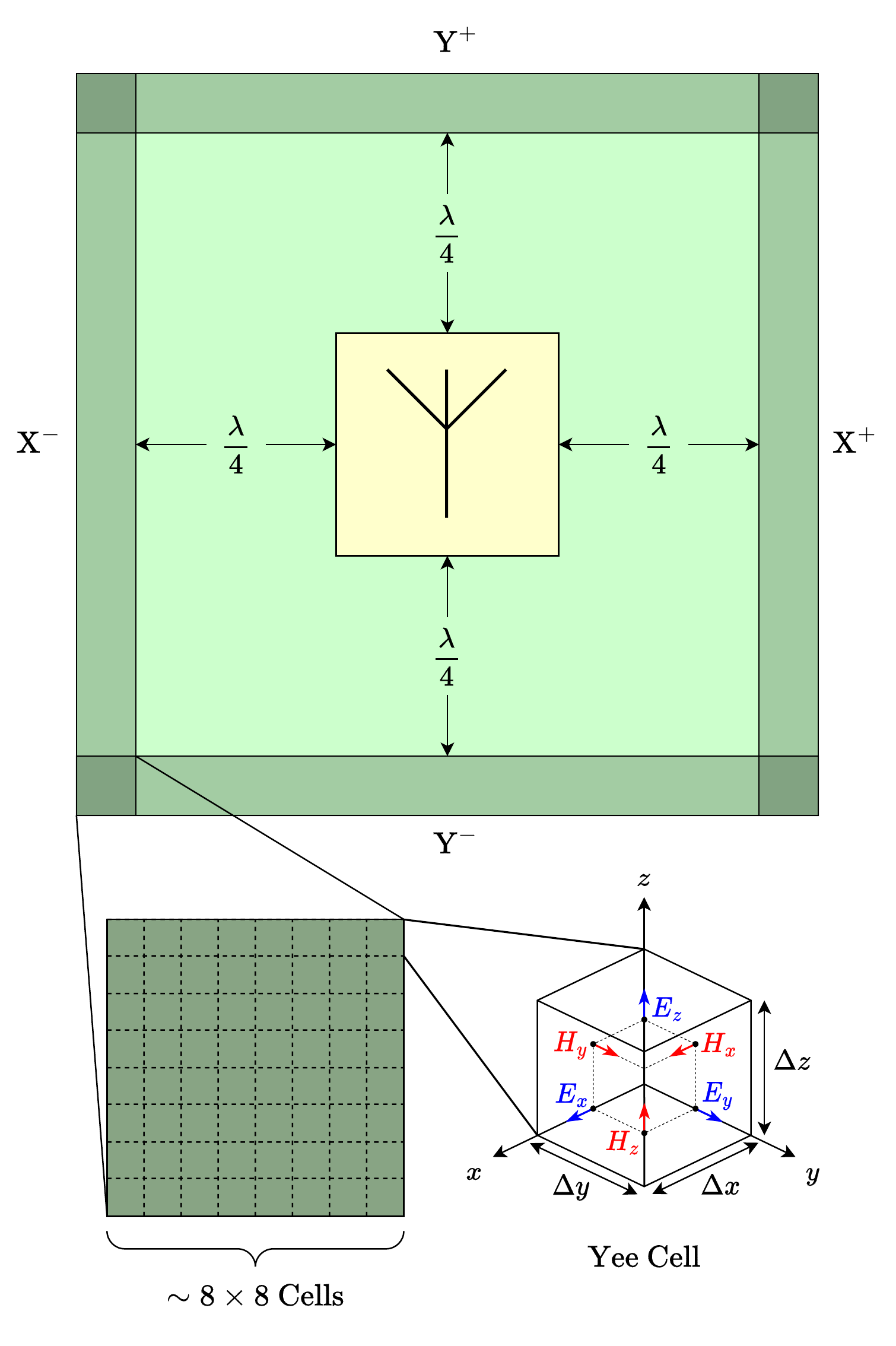}
  \caption{\textit{Top:} Two-dimensional view of the geometry placed the simulation domain. The PML zone (meant to absorb incident energy) is shaded, and is separated from the edges of the model by a quarter of the wavelength. \textit{Bottom left:} Local grid structure of the PML region, highlighting its composition of multiple cells. \textit{Bottom right:} The three-dimensional Yee cell structure of each mesh element, upon which the fundamental operation of the FDTD method is based.}\label{fig:pml}
  \end{center}
\end{figure}

\begin{figure*}
  \begin{center}
  \includegraphics[width=\textwidth]{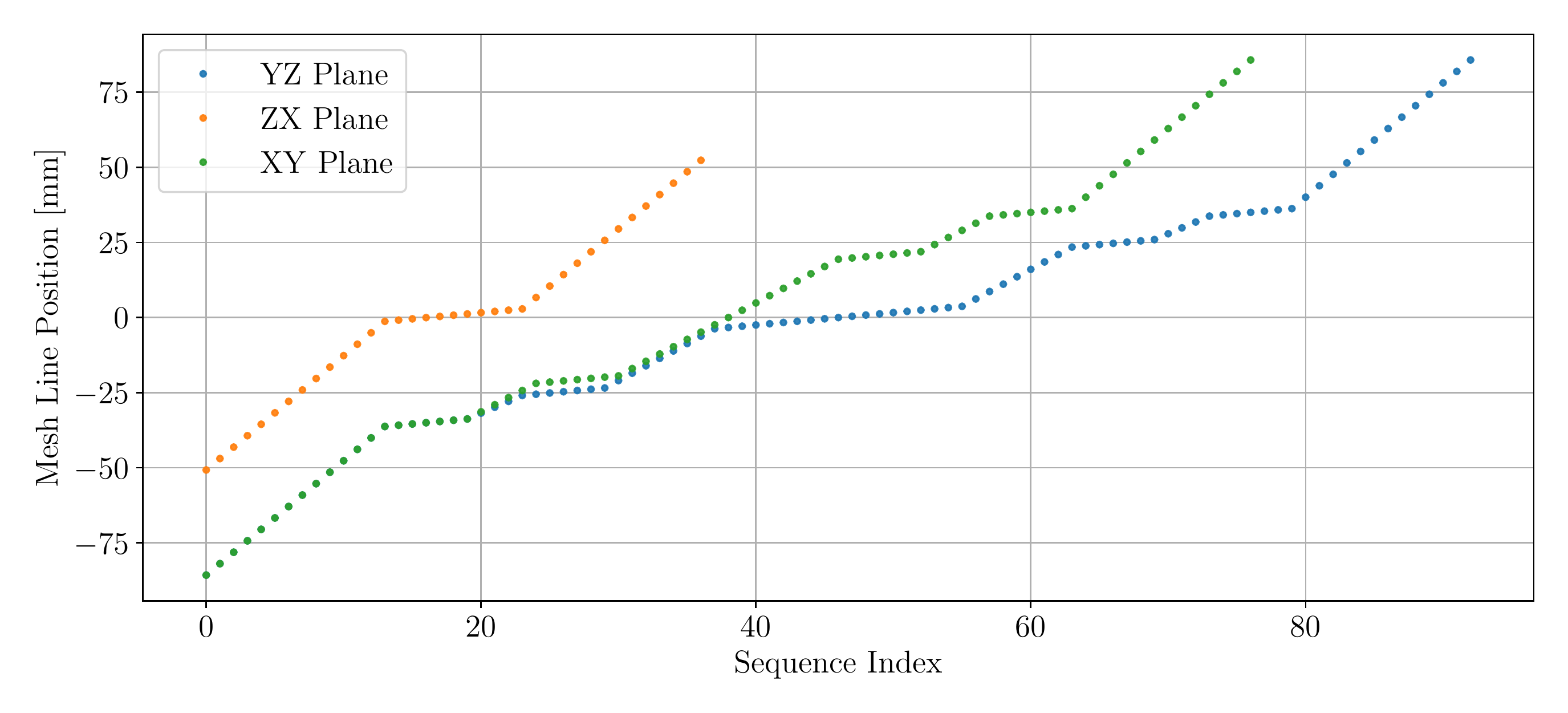}
  \caption{Mesh sequence for each plane in $\mathbb{R}^3$, consisting of $93\times37\times77=264{,}957$ mesh cells in total. Note the decreasing slope around every $m$, as the concentration of mesh lines becomes denser wherever refinement is deemed necessary by the algorithm.}\label{fig:sequence}
  \end{center}
\end{figure*}

\section{Algorithm Inputs}
In order for the algorithm to properly generate a grid suitable for the model, three inputs are required:

\begin{enumerate}
   \item[{1.}] \texttt{max\_cell\_model} \ \ \ \textit{(float)}
   \item[{2.}] \texttt{max\_cell\_space} \ \ \  \textit{(float)}
   \item[{3.}] \texttt{min\_cell\_global}  \ \textit{(float)}\\
\end{enumerate} Additionally, two optional parameters can assist with further refinements of the model for more advanced control:

\begin{enumerate}
   \item[{4.}] \texttt{n} \ \ \ \ \ \ \ \ \ \ \ \ \ \ \ \ \ \ \ \ \ \ \ \textit{(list)}
   \item[{5.}] \texttt{res\_fraction} \ \ \ \ \ \ \textit{(list)}\\
\end{enumerate} Inputs 1--3 are expressed as fractions of the shortest wavelength $\lambda_\mathrm{min}$ the excitation signal consists of. In time-domain solvers, this is determined by the maximum frequency of the Fourier transform of the excitation signal. Thus, if a generated grid is sufficiently fine for the highest frequency of the excitation, it is inherently applicable to all lower frequencies, given

\begin{equation}\label{eq:wavelength}
\frac{c}{\lambda_\mathrm{min}} > \frac{c}{\lambda} > \frac{c}{\lambda_\mathrm{max}}, \ \ \ \forall \lambda \in (\lambda_\mathrm{min}, \lambda_\mathrm{max})\\
\end{equation} where $c$ is the speed of light. However, although the level of detail may suffice, it is important to consider the space between the end of the model to the boundaries of the simulation, which are partially dependant on the longest wavelength of the excitation.\\

The distance between the edge of the geometry to the absorbing boundary condition (ABC) is generally given by the midpoint of the two wavelengths, $\lambda_\mathrm{mid}$, and is typically set to a quarter of the wavelength:

\begin{equation}\label{eq:wavelength_mid}
\frac{1}{4} \lambda_\mathrm{mid} = \frac{\lambda_\mathrm{min}\cdot \lambda_\mathrm{max}}{2(\lambda_\mathrm{min}+\lambda_\mathrm{max})}. \\
\end{equation} This ensures enough cells are provided for the fields to form to an acceptable degree, before reaching the ABC (and go through e.g., nearfield-to-farfield transformations).\\

While it is theoretically more appropriate to use a quarter of the longest wavelength $\lambda_\mathrm{max}$ instead, this becomes highly impractical when the excitation consists of very low frequencies, as the number of mesh cells can grow tremendously due to the additional space that needs to be meshed. Moreover, the simulation domain would become infinitely large if the excitation signal included a frequency $f$ equal to $0 \mathrm{\ Hz}$ (i.e., not be DC--free):

\begin{equation}\label{eq:infinite_bounds}
\lim_{f \to 0^{+}} \frac{1}{4} \lambda_\mathrm{max} = \infty.\\
\end{equation}

\newcommand{\minus}{\scalebox{0.8}{$-$}}
\newcommand{\plus}{\scalebox{0.6}{$+$}}

Additionally, depending on the type of boundaries set for the simulation, setting a quarter of the wavelength as the distance between the end of the geometry to the simulation domain may not be appropriate. For instance, because the cells of the perfectly matched layer (PML) extend inward of the boundaries, the radiated fields from the source may not have enough space to fully form, ultimately getting absorbed by PML cells prematurely. In certain applications, such setup imperfections may potentially yield inaccurate simulation results. For that reason, $\sim8$ additional cells (optionally adjustable \texttt{pml\_n} \textit{integer}, $n\in [\![4,50]\!]$) are appended to the six ends of the quarter wavelength spacings $(\mathrm{X}^{-}, \mathrm{X}^{+}, \mathrm{Y}^{-}, \mathrm{Y}^{+}, \mathrm{Z}^{-}, \mathrm{Z}^{+})$. Fig.~\ref{fig:pml} depicts this concept in two dimensions $(\mathrm{X}, \mathrm{Y})$.

\begin{figure}
  \begin{center}
  \includegraphics[width=0.44\textwidth]{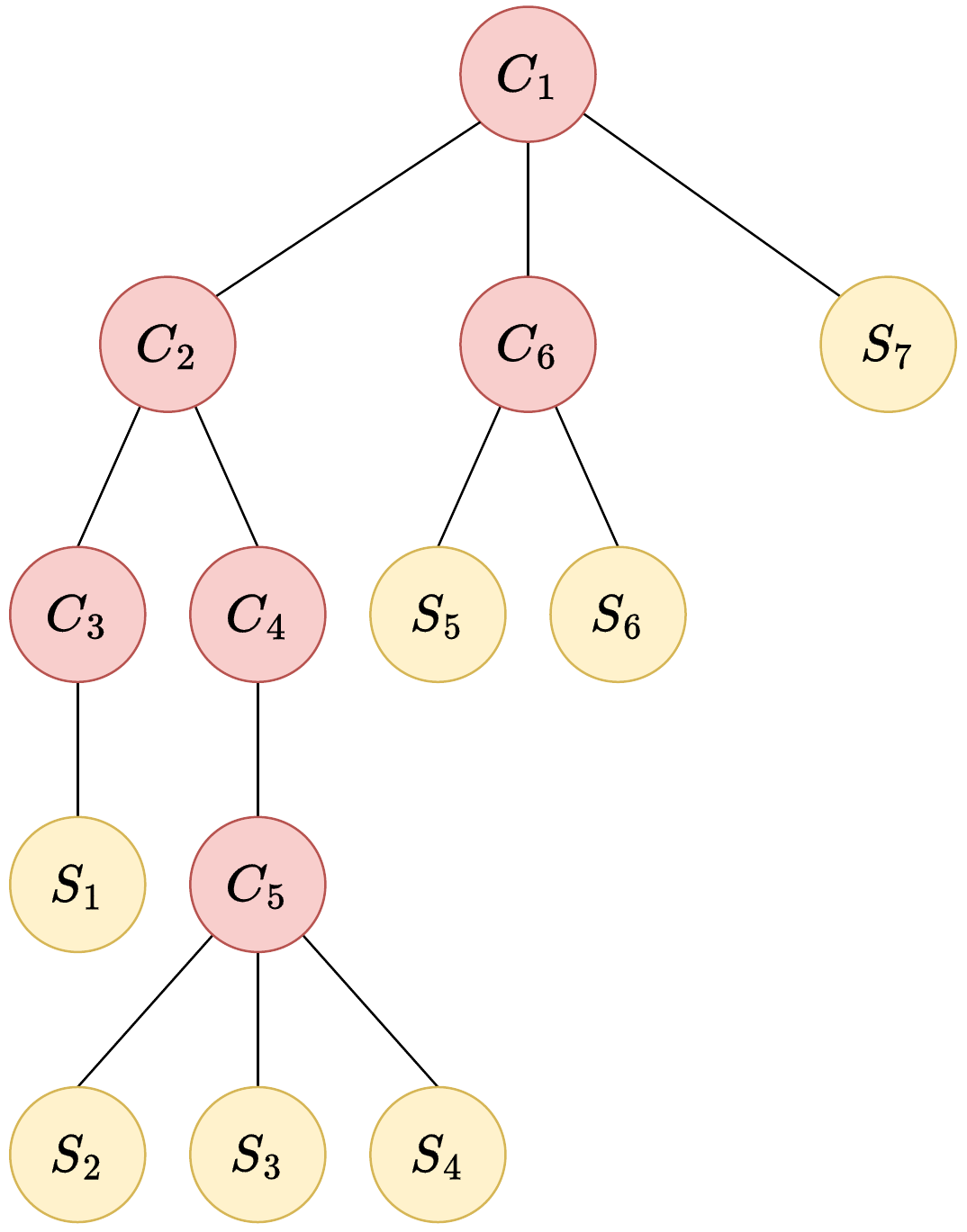}
  \caption{Tree data structure of a sample geometry model. $C_i$ and $S_j$ correspond to the $i$-th compound and $j$-th shape of the input file, respectively. Assuming the root node to be $C_1$, the output of the DFS algorithm in this case would be $C_1$,~$C_2$,~$C_3$,~$S_1$,~$C_4$,~$C_5$,~$S_2$,~$S_3$,~$S_4$,~$C_6$,~$S_5$,~$S_6$,~$S_7$. However, since we are only interested in shapes, we can eliminate all compound instances, yielding the final result of $S_1$,~$S_2$,~$\cdots$,~$S_7$.}\label{fig:dfs}
  \end{center}
\end{figure}

\subsection{Significant Parameters}

The \texttt{max\_cell\_model} parameter is the primary variable that specifies the local grid resolution of the model being simulated. The greater the value, the finer the produced mesh around the corresponding region.\\

Similar to this parameter, \texttt{max\_cell\_space} describes the grid resolution of the space between the ends of the geometry to the bounding box of the simulation domain. In the presence of an absorbing PML, the size of these cells are also affected accordingly. Because of the structural complexity of the model, due to abrupt differences in permittivity, $\Delta \varepsilon$, and permeability, $\Delta \mu$, the response of the electromagnetic fields tends to be highly variable and sudden. On the other hand, since the free-space region surrounding the model has a constant vacuum permittivity $\varepsilon_0$ and permeability $\mu_0$, the electromagnetic variations the waves exhibit as they propagate toward the PML zones are minimal.\\
\ \\
\ \\
\ \\
Thus, most cases are recommended to follow:

\begin{equation}\label{eq:model_vs_space}
\texttt{max\_cell\_model} > \texttt{max\_cell\_space}.\\
\end{equation}

Furthermore, the \texttt{min\_cell\_global} parameter determines the minimum cell dimension for all axes. Unlike the aforementioned parameters which are either only applied to the geometry, or the surrounding free-space region individually, the value of this parameter ensures no cell exceeds a specified wavelength fraction, globally across the entire simulation domain. Imposing such a constraint to the grid is important for two main reasons; namely, the number of elements in the mesh can be kept low, even if the model suffers from areas with extremely fine detail, and secondly, the FDTD timestep can be kept long. Recall the simulation stability condition imposed by the Courant--Friedrichs--Lewy (CFL) condition \cite{cfl, cfl_yee}:

\begin{equation}\label{eq:cfl}
\Delta t \leq \frac{\displaystyle 1}{\displaystyle c\sqrt{\frac{1}{{\Delta x}^2}+\frac{1}{{\Delta y}^2}+\frac{1}{{\Delta z}^2}}},\\
\end{equation}
where $\Delta t$ is the timestep imposed by the constraint, $c$ is the speed of light, and $\Delta x$, $\Delta y$ and $\Delta z$ correspond to the minimum cell edge length in the $x$, $y$ and $z$ axis, respectively. Therefore, by ensuring no cell exceeds a specified fraction, such as $\lambda/300$, tiny details in the geometry of the model---including e.g.,~thin PCB and microstrip traces with a thickness of $\sim35${\ \textmu}m---which fail to contribute anything to the model can be eliminated, as they are electromagnetically unnecessary. This in turn minimizes the total number of timesteps $N_\mathrm{ts}$, ultimately leading to a faster simulation; often with a considerable speedup factor.

\subsection{Optional Parameters}

As mentioned, sudden changes in the material properties between two neighboring mesh cells can have a meaningful impact on the output of a simulation. Let $(\varepsilon_n)$ and $(\mu_n)$ be the sequences of an axis $\in \mathbb{R}^3$, indicating the permittivity and permeability at the $n$-th cell of the axis array, respectively.\footnote{Unlike the sequence describing the position of each mesh line which is strictly increasing, $(\varepsilon_n)$ and $(\mu_n)$ are typically non-injective, as the same material is generally part of several cells along the same axis.} If we consider that $\exists\ m\in\mathbb{N}$, such that

\begin{equation}\label{eq:and_or}
\begin{aligned}
\varepsilon_{m+1} - \varepsilon_m \neq 0\ \ \ &\lor\ \ \ \mu_{m+1} - \mu_m \neq 0\\
\Delta \varepsilon \neq 0\ \ \ &\lor\ \ \ \Delta \mu \neq 0,
\\
\end{aligned}
\end{equation}
it becomes apparent that it is of great importance to ensure the mesh is appropriately refined $\forall m$ for which Eq. (\ref{eq:and_or}) holds true. I.e., for every edge around which such variations are expected, additional lines shall be inserted to assure a smooth transition between materials of different properties. If we assume a material-based boolean operation (addition; shape fuse)\footnote{\url{https://cadquery.readthedocs.io/en/latest/classreference.html\#cadquery.Compound.fuse}} has preceded prior to the geometry being fed to the core segment of the mesh generation algorithm, such that $N_\mathrm{shapes} = N_\mathrm{materials}$, the number of edges and vertices will be minimized. By refining each vertex of the CAD model, we can therefore practically ensure all shape edges are properly refined. Fig.~\ref{fig:sequence} demonstrates edge refinement on a sample microstrip patch antenna case.\\

By default, edge refinement places 3 additional lines around each side of the corresponding node, symmetrically (yielding a total of 6 additional lines for every $m$). In case the model could benefit from certain special tweaking, this can be optionally adjusted using the \texttt{n} variable, which allows the constant to be configured to a lower or higher value for each plane, individually.\\

Similarly, the \texttt{res\_fraction} parameter list can adjust the wavelength fraction for each plane ($\lambda/6$ by default), so that finer details can be picked up, or for the mesh to become further adapted to maximize the timestep $\Delta t$ (CFL constraint; Eq. (\ref{eq:cfl})).

\begin{figure*}
  \begin{center}
  \includegraphics[width=\textwidth]{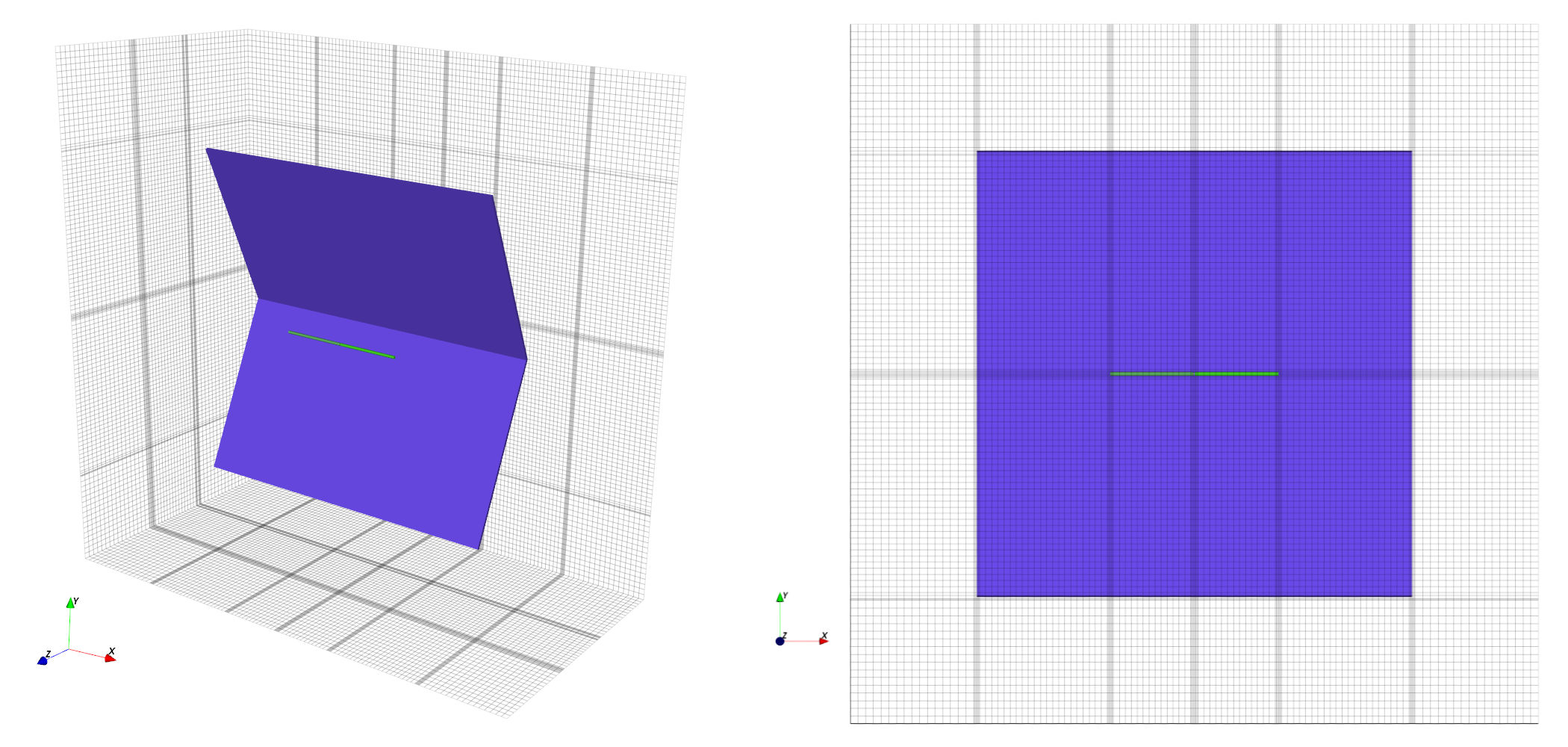}
  \caption{Produced mesh generated using the proposed algorithm on a dipole-fed corner reflector antenna. \textit{Left:} Three-dimensional view of the derived grid. \textit{Right:} Two-dimensional parallel projection of the XY plane, depicting the refined regions and the sparsely-separated grid lines far from the model. The grid consists of $132\times118\times64=996{,}864$ mesh cells and was produced in under 2 seconds on an Intel Core i7-10750H CPU @ 2.60 GHz ($<$1 s for CAD preprocessing and $<$0.3 s for mesh generation).}\label{fig:wire_mesh}
  \end{center}
\end{figure*}

\section{Depth-First Search}

Because the algorithm expects a STEP file as an input (to support a broad range of CAD geometries in a compatible manner), it is important for the individual shapes to be separated from compounds (groups of shapes). However, due to the variety of designs imposing different CAD formats, the expected structure of the relationship between compounds and shapes is unknown. For that reason, we choose to interpret the model's structure as a tree consisting of an arbitrary number of nodes, depending on the input model. The internal nodes represent the compounds, while the leaf nodes represent the shapes of interest we wish to obtain.\\

While CAD preprocessing is not strictly considered to be a part of the core mesh-generation algorithm, it unquestionably constitutes a critical step required for the derivation of the geometry vertices of the input model. Handling the geometry thus becomes a node searching problem, where a tree data structure needs to be traversed to obtain each individual shape the geometry consists of.\\

In order to traverse the tree, we can employ the depth-first search (DFS) algorithm (\cite{dfs}). Beginning from a root node, DFS works by exploring each branch individually, ultimately retrieving all leaf nodes of interest. This makes DFS an ideal candidate for retrieving the shapes from compounds, as certain compounds may contain several other compounds (arranged in the form of a subtree), before unveiling its shape(s). In other words, nested compounds are possible and expected. Fig.~\ref{fig:dfs} shows an example geometry input, translated into its tree data structure.

\section{Accuracy Evaluation}

\begin{figure*}
  \begin{center}
  \includegraphics[width=\textwidth]{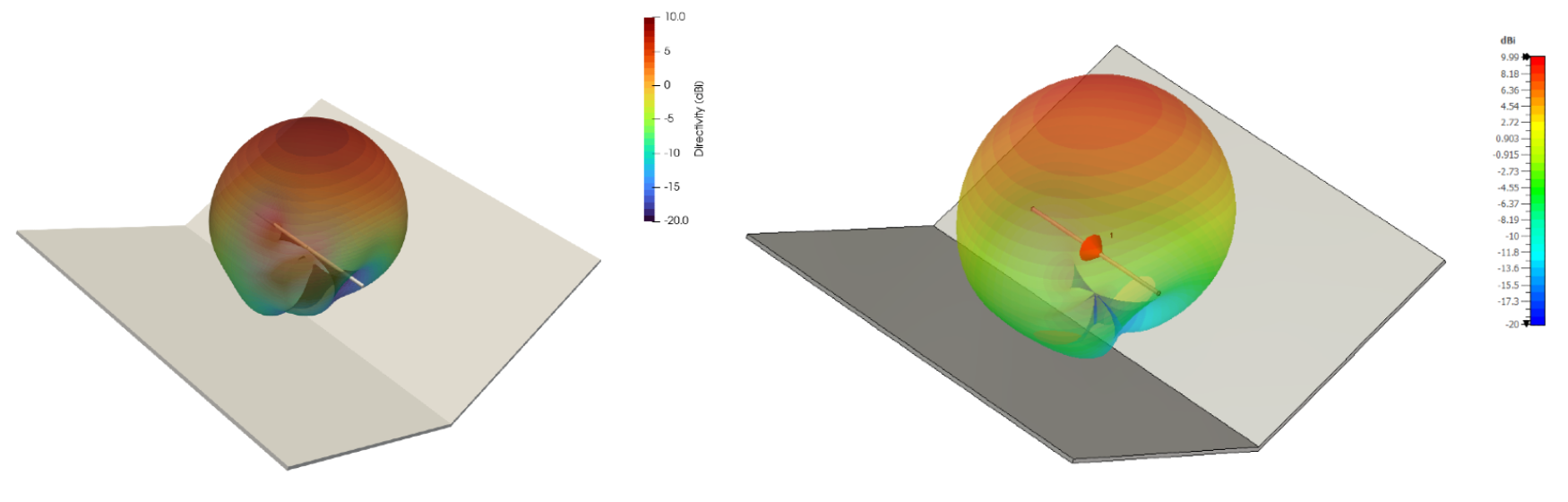}
  \includegraphics[width=\textwidth]{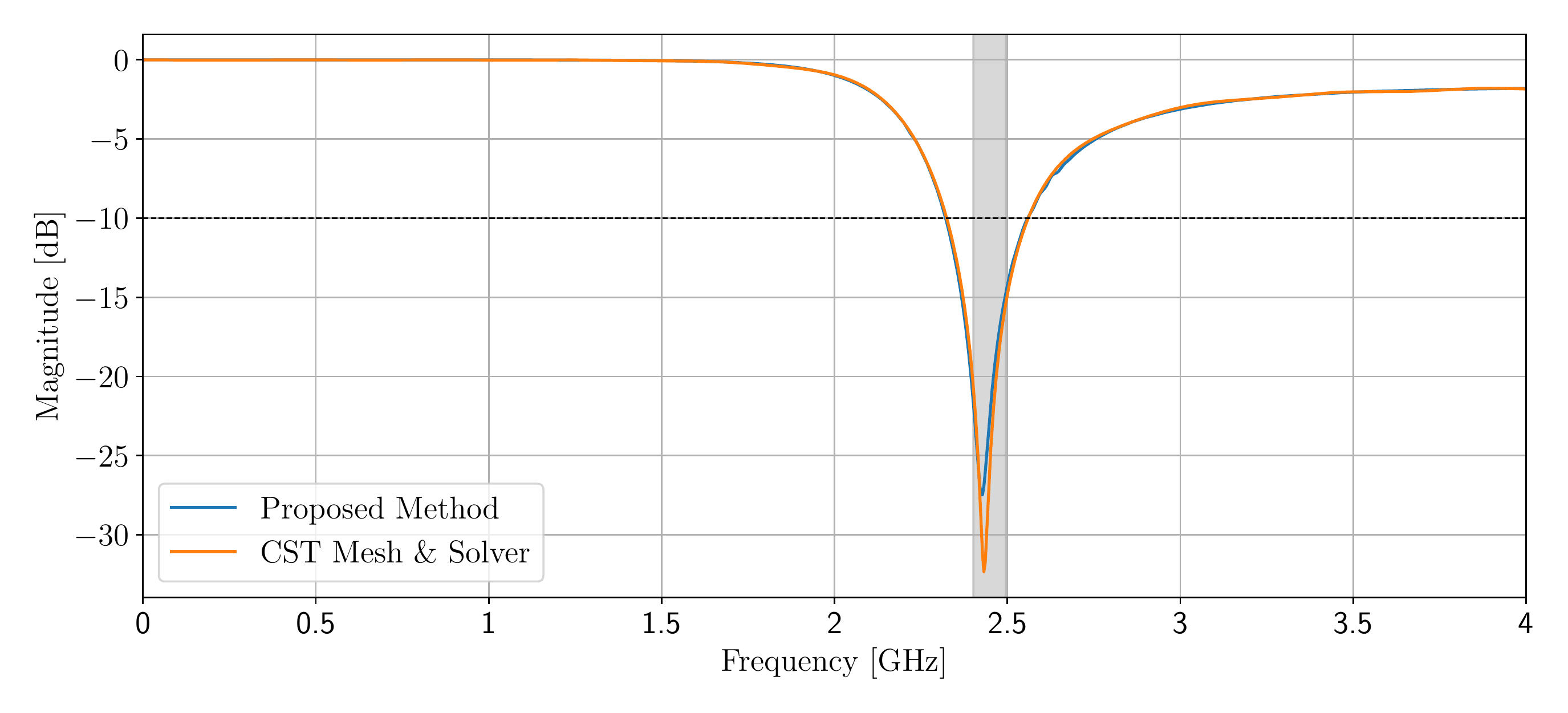}
  \caption{Results derived using the proposed meshing algorithm compared to CST's proprietary mesher and time-domain solver. \textit{Top:} 3D farfield at 2.4 GHz, showing the radiation pattern produced using our method (left) and CST (right). The maximum boresight directivity derived using EC--FDTD and CST's solver is 9.81 and 9.99 dBi, respectively. \textit{Bottom:} Comparison of the reflection coefficient $(|\mathrm{S}_{11}|)$ between the two simulations, showing solid agreement across the entire DC---4 GHz range. The shaded region (2,401---2,495 MHz) corresponds to the 14 channels of the IEEE 802.11b/g standard for wireless local area network (WLAN) communications in the 2.4 GHz band.}\label{fig:wire_results}
  \end{center}
\end{figure*}

In order to evaluate the robustness and quality of the proposed algorithm, the simulation results produced with the generated grids are compared with those of commercial tools. Considering its similarities in terms of mesh generation and time-domain approach to Maxwell's equations, CST Studio Suite has been used as a baseline for our results. In Fig.~\ref{fig:wire_mesh}, we show the mesh produced using the following parameters for a dipole-fed corner reflector antenna designed for Wi-Fi applications:

\begin{itemize}
   \item \texttt{max\_cell\_model} = 40
   \item \texttt{max\_cell\_space} = 30
   \item \texttt{min\_cell\_global} = 300\\
\end{itemize} The simulation has been carried out using the open-source equivalent-circuit finite-difference time-domain (EC--FDTD) solver offered by the openEMS tool (\cite{openems_2}). The results (Fig.~\ref{fig:wire_results}) are in agreement with CST in terms of the reflection coefficient, pattern shape, as well as peak directivity.\footnote{The results are also in solid agreement in terms of the time-domain voltage response and $\mathrm{S}_{11}$ phase, but were not deemed meaningfully important to include as separate figures.}

\section{Conclusions}

Despite the numerous advancements in computational electromagnetics over the last decades, no reliable open-source solution had been available for a stage as critical as geometry discretization. While numerous tetrahedral meshing algorithms have existed for many years due to the broad range of applications found in the various fields of simulation engineering, a hexahedral rectilinear mesher---applicable to the popular finite-difference time-domain method---had not been available.\\

We have presented a robust and automatic approach to tackle this problem in a highly efficient and computationally inexpensive manner, that enables the algorithm to be used by antenna designers and RF engineers to easily mesh their complex geometries in a highly automated and greatly simplified manner. Furthermore, we have compared our results with commercial software, and have demonstrated clear agreement between both the produced grid, as well as the simulation output derived with each mesh and solver.\\

In future work, we hope to develop an accurate binary classifier based on machine learning, with the aim of introducing a highly intelligent system capable of determining whether a three-dimensional non-uniform structured grid is fine enough, based on mesh convergence. Assuming the training dataset of antenna models with varying mesh resolutions is sufficiently large, such a neural network is expected to drastically reduce the number of mesh cells required to produce an accurate result (using image recognition), leading to a significantly faster simulation. Furthermore, if the false positive rate of the classifier is found to be exceptionally low, conventional mesh convergence analysis could potentially become redundant.\\

Additionally, considering the openEMS package supports an EC--FDTD implementation in cylindrical coordinates as well, a similar algorithm could be adjusted to be applicable and adaptable to geometries that present curvature, where the FDTD staircase representation problem induced by conventional Cartesian grids could be avoided.

\chapter{OptimizeRF: A Modular Optimization Interface for Network Analyzers}
The optimization of RF networks is usually carried out on a simulation level, where parameters are tuned in software and the results are derived using numerical methods. However, the minimization/maximization of objective functions associated with certain performance metrics of networks has several applications in the physical world, such as mass prototyping and reconfigurability.\\

Furthermore, it is often significantly faster to physically tune parameters using e.g. stepper motors, because the evaluation of the objective function is simply an instrument measurement. On the contrary, simulations require the entire system to be simulated, which can take up to several hours to get the objective function to return a value, depending on the size of the electromagnetic problem.\\

\begin{figure}[htbp]
\centering
\includegraphics[width=1\textwidth]{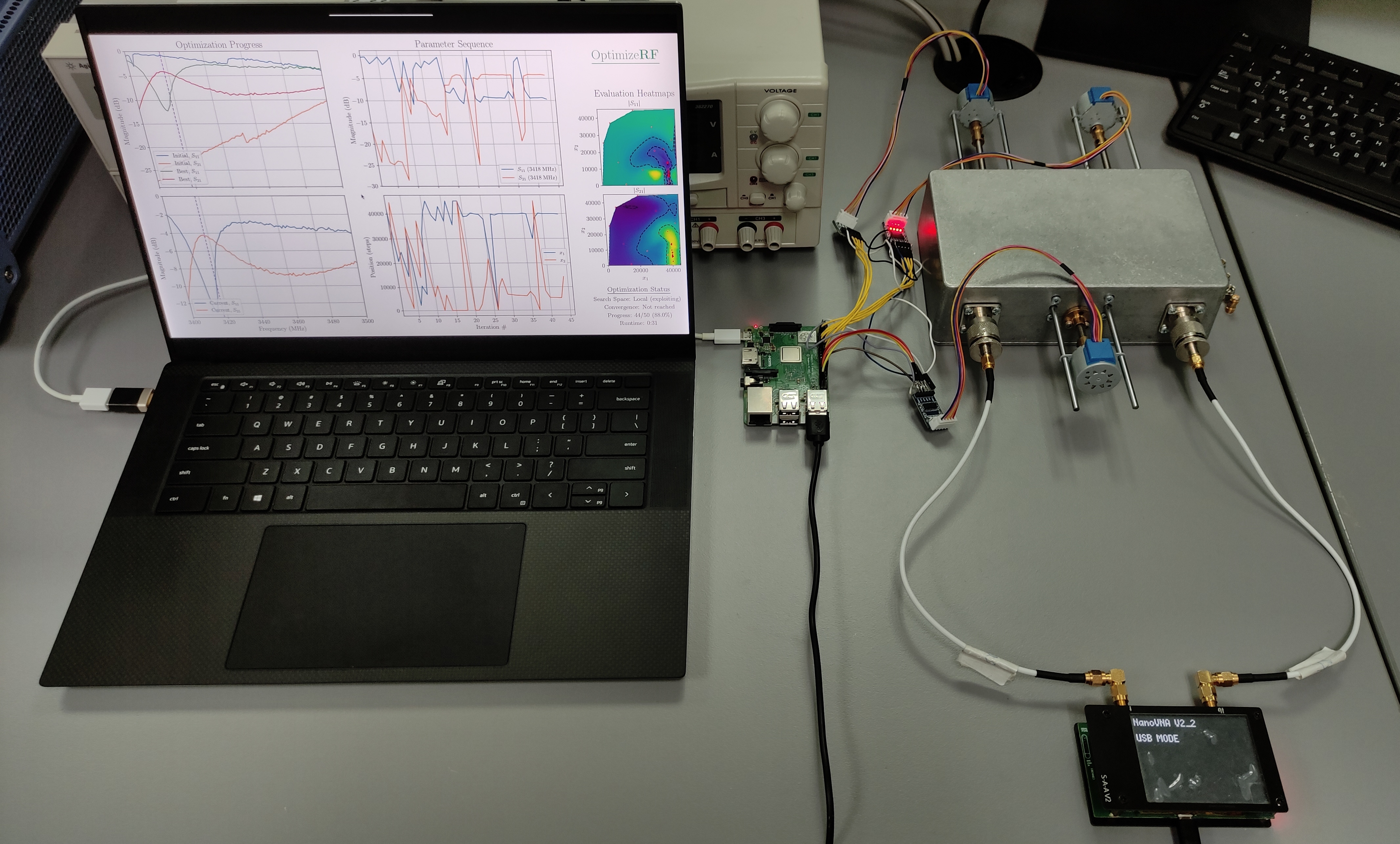}

\includegraphics[width=1\textwidth]{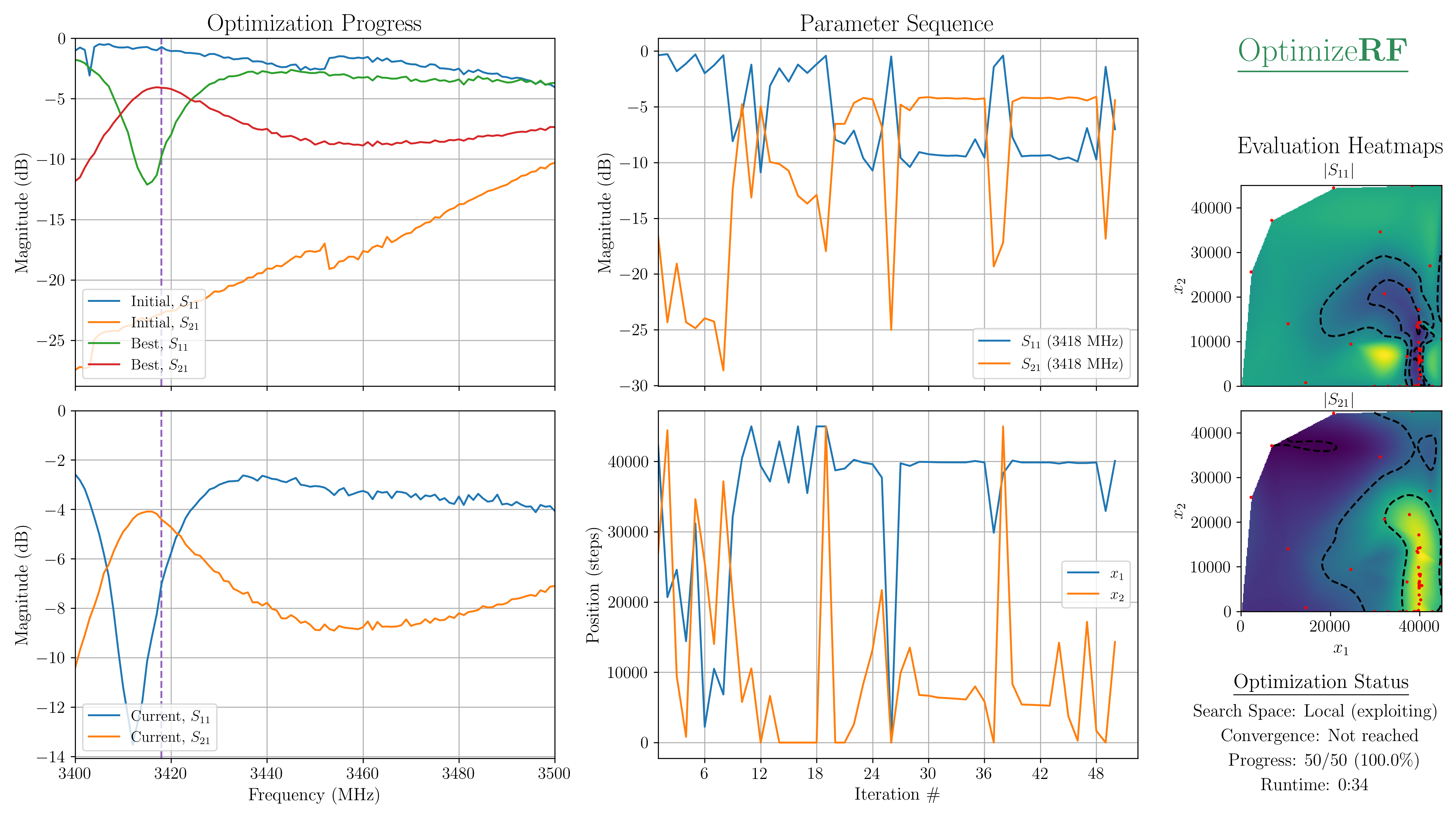}

\caption{\textit{Top:} Setup including the network/device under test (DUT) whose scattering parameter $(|\mathrm{S}_{21}|)$ requires optimization (maximization at 3,418 MHz), a low-cost vector network analyzer (VNA), and a Raspberry Pi for the acquisition and processing of the data, as well as the decision-maker of the following iterations, based on the optimization algorithm. The stepper motors are connected via three motor drivers. \textit{Bottom:} Results of a 2-parameter optimization.}
\label{fig:optimizerf}
\end{figure}

In the Appendix, the source code of OptimizeRF is provided: a modular optimization interface for network analyzers, with a proof-of-concept demonstration on a custom-built band-pass interdigital cavity filter with a center frequency of 3.42 GHz (Fig.~\ref{fig:optimizerf}). Considering the transfer function/magnitude of scattering parameters $(|\mathrm{S}_{ij}|)$ can be set as the objective function, the same methodology is applicable to the optimization of reconfigurable intelligent surfaces. The optimizer supports real values (floats), with (optionally) different bounds for each parameter.\\

The optimization algorithm is based on the \texttt{gp\_minimize}\footnote{\url{https://scikit-optimize.github.io/stable/modules/generated/skopt.gp_minimize.html}} function provided by the \texttt{scikit-optimize} Python package (\cite{skopt}). Unlike conventional optimization methods used in the field of simulation engineering, OptimizeRF is based on Bayesian optimization using Gaussian Processes: an optimization algorithm suitable for noisy objective functions, such as those returned by VNAs.\\

This work is currently under preparation for publication.\\

\begin{figure}[htbp]
\centering
\includegraphics[width=1.1\textwidth]{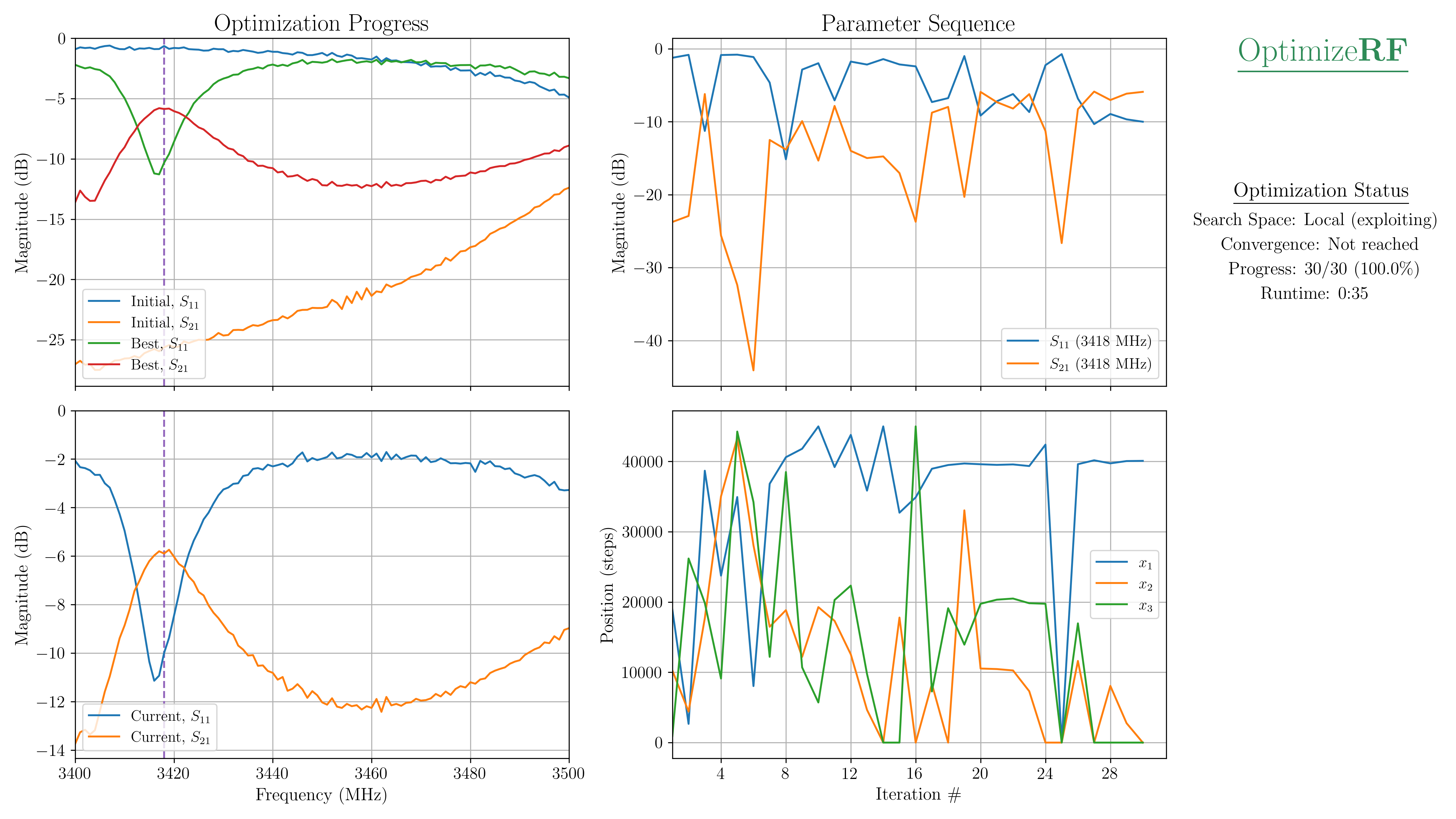}
\caption{Results of a 3-parameter optimization (all tuning rods in motion). The result is in agreement with the 2-parameter optimization, implying minimal sensitivity on 1 out of 3 parameters on the response (transfer function) of the device.}
\label{fig:optimizerf_3params}
\end{figure}



\phantomsection
\printbibliography 
\addcontentsline{toc}{chapter}{Bibliography}



\appendix
\chapter{Source Code}



\section{OptimizeRF Module Code}
\begin{lstlisting}
#!/usr/bin/python
# -*- coding: utf-8 -*-

from datetime import datetime
import argparse
from nanovna import NanoVNAV2
import skrf
from skrf.media import Coaxial
import numpy as np
import matplotlib.pyplot as plt
from matplotlib.ticker import MaxNLocator
from matplotlib import gridspec
import time
from datetime import timedelta
import threading
import os
from skopt import gp_minimize
from scipy import interpolate
import ast
from tkinter import *
import sys

### Matplotlib configuration ###

default_params = {
    'toolbar': 'None',
    'figure.dpi': 120,
    'figure.figsize': [4, 3],
    'figure.subplot.left': 0.15,
    'figure.subplot.right': 0.9,
    'figure.subplot.bottom': 0.12,
    'axes.titlesize': 'medium',
    'axes.labelsize': 14,
    'ytick.labelsize': 'small',
    'xtick.labelsize': 'small',
    'legend.fontsize': 12,
    'legend.loc': 'best',
    'font.size': 16,
    'font.family': 'serif',
    'text.usetex': True,
    }
plt.rcParams.update(default_params)


def optimize(
    f_start,
    f_stop,
    sweep_pts,
    cal,
    load_cal,
    f_goal,
    goal,
    s_parameter,
    parameters,
    bounds,
    reset,
    max_calls,
    exp_ratio,
    conv_ratio,
    output,
    ):
    global start_time, positions, best, ntw_best, s, grid_x, grid_y, \
        s11_progress, s21_progress, f_goal_label, flag

    # ## Input parameters ###
    # Goals

    if goal.lower() == 'min' or goal.lower() == 'minimize':
        sign = 1  # Minimize S-Parameter
    elif goal.lower() == 'max' or goal.lower() == 'maximize':
        sign = -1  # Maximize S-Parameter

    # Import parameters

    modules = []
    for parameter in parameters:
        try:
            modules.append(__import__(parameter))
        except ImportError:
            print ('Error: Failed to import \'', parameter, '\'.')
            return 1

    # Parameter bounds

    bounds = ast.literal_eval(bounds)

    f_arr = np.linspace(f_start / 1e6, f_stop / 1e6, num=sweep_pts)
    f = skrf.Frequency.from_f(f_arr, unit='MHz')

    # ## Clear previous data/logs ###

    files = ['Current.s2p', 'Initial.s2p', 'Best.s2p', 'log.txt']

    for file in files:
        if os.path.isfile(file):
            os.remove(file)
        else:
            pass

    # ## Device initiation ###
    # Initiate device

    nv = NanoVNAV2()

    # Set frequency range

    nv.set_sweep(f_start, f_stop)
    nv.fetch_frequencies()

    calibrated = False
    if load_cal != '':
        calibrated = True
    else:
        load_cal = '.'
        if cal:

            # print('Performing measurement calibration...\n')

            with open('log.txt', 'a') as log:
                log.write('===============================\n')
                log.write('[*] Performing measurement calibration...\n')

            # ## OPEN ###

            input('Connect OPEN to Port 1 and press Enter.')

            # Measure S11 and S21

            s11 = nv.data(0)
            s21 = nv.data(1)

            f_arr = np.linspace(f_start / 1e6, f_stop / 1e6,
                                num=sweep_pts)
            f = skrf.Frequency.from_f(f_arr, unit='MHz')
            s = np.zeros((len(f), 2, 2), dtype='complex_')
            s[:, 0, 0] = s11
            s[:, 0, 1] = s21  # # np.zeros_like(s11) #np.ones_like(s11)*np.nan
            s[:, 1, 0] = s21
            s[:, 1, 1] = s11  # # np.ones_like(s11) #np.ones_like(s11)*np.nan
            ntw_init = skrf.Network(frequency=f, s=s)

            ntw_init.write_touchstone(load_cal + '/open',
                    skrf_comment=False)

            # ## SHORT ###

            input('Connect SHORT to Port 1 and press Enter.')

            # Measure S11 and S21

            s11 = nv.data(0)
            s21 = nv.data(1)

            f_arr = np.linspace(f_start / 1e6, f_stop / 1e6,
                                num=sweep_pts)
            f = skrf.Frequency.from_f(f_arr, unit='MHz')
            s = np.zeros((len(f), 2, 2), dtype='complex_')
            s[:, 0, 0] = s11
            s[:, 0, 1] = s21  # # np.zeros_like(s11) #np.ones_like(s11)*np.nan
            s[:, 1, 0] = s21
            s[:, 1, 1] = s11  # # np.zeros_like(s11) #np.ones_like(s11)*np.nan
            ntw_init = skrf.Network(frequency=f, s=s)

            ntw_init.write_touchstone(load_cal + '/short',
                    skrf_comment=False)

            # ## LOAD ###

            input('Connect LOAD to Port 1 and press Enter.')

            # Measure S11 and S21

            s11 = nv.data(0)
            s21 = nv.data(1)

            f_arr = np.linspace(f_start / 1e6, f_stop / 1e6,
                                num=sweep_pts)
            f = skrf.Frequency.from_f(f_arr, unit='MHz')
            s = np.zeros((len(f), 2, 2), dtype='complex_')
            s[:, 0, 0] = s11
            s[:, 0, 1] = s21  # # np.zeros_like(s11) #np.ones_like(s11)*np.nan
            s[:, 1, 0] = s21
            s[:, 1, 1] = s11  # # np.ones_like(s11) #np.ones_like(s11)*np.nan
            ntw_init = skrf.Network(frequency=f, s=s)

            ntw_init.write_touchstone(load_cal + '/load',
                    skrf_comment=False)

            # ## THRU ###

            input('Connect THRU between Port 1 & 2 and press Enter.')

            # Measure S11 and S21

            s11 = nv.data(0)
            s21 = nv.data(1)

            f_arr = np.linspace(f_start / 1e6, f_stop / 1e6,
                                num=sweep_pts)
            f = skrf.Frequency.from_f(f_arr, unit='MHz')
            s = np.zeros((len(f), 2, 2), dtype='complex_')
            s[:, 0, 0] = s11
            s[:, 0, 1] = s21  # # np.ones_like(s11) #np.ones_like(s11)*np.nan
            s[:, 1, 0] = s21
            s[:, 1, 1] = s11  # # np.zeros_like(s11) #np.ones_like(s11)*np.nan
            ntw_init = skrf.Network(frequency=f, s=s)

            ntw_init.write_touchstone(load_cal + '/thru',
                    skrf_comment=False)
            calibrated = True
    if calibrated:

        # ## Create ideals ###

        coax = Coaxial(frequency=f, z0=50)
        my_ideals = [coax.short(nports=2), coax.open(nports=2),
                     coax.match(nports=2), coax.thru()]
        my_measured = [skrf.Network(load_cal + '/short.s2p'),
                       skrf.Network(load_cal + '/open.s2p'),
                       skrf.Network(load_cal + '/load.s2p'),
                       skrf.Network(load_cal + '/thru.s2p')]

        # # Create TwoPortOnePath instance

        calibration = \
            skrf.calibration.TwoPortOnePath(measured=my_measured,
                ideals=my_ideals)

        # Run calibration algorithm

        calibration.run()
        if cal:
            input('Calibration complete. Press Enter to begin DUT optimization.'
                  )

    # Measure initial S11 and S21

    s11 = nv.data(0)
    s21 = nv.data(1)

    f_arr = np.linspace(f_start / 1e6, f_stop / 1e6, num=sweep_pts)
    f = skrf.Frequency.from_f(f_arr, unit='MHz')
    s = np.zeros((len(f), 2, 2), dtype='complex_')
    s[:, 0, 0] = s11
    s[:, 0, 1] = s21  # # np.ones_like(s11)*np.nan #np.zeros_like(s11)
    s[:, 1, 0] = s21
    s[:, 1, 1] = s11  # # np.ones_like(s11)*np.nan #np.ones_like(s11)
    ntw_init = skrf.Network(frequency=f, s=s)
    if calibrated:
        ntw_init = calibration.apply_cal((ntw_init, ntw_init))

    s = ntw_init.s

    ntw_init.write_touchstone('Initial', skrf_comment=False)
    ntw_init = skrf.Network('Initial.s2p')

    ntw_best = ntw_init
    ntw_best.write_touchstone('Best', skrf_comment=False)
    ntw_best = skrf.Network('Best.s2p')

    f_goal /= 1e6

    idx = np.abs(f_arr - f_goal).argmin()

    if f_goal == int(f_goal):
        f_goal_label = str(int(f_goal))
    else:
        f_goal_label = str(f_goal)

    positions = np.empty((0, len(bounds)))
    s11_progress = np.array([])
    s21_progress = np.array([])
    if len(bounds) == 2:
        (grid_x, grid_y) = np.mgrid[bounds[0][0]:bounds[0][1]:250j,
                                    bounds[1][0]:bounds[1][1]:250j]

    plt.ion()

    fig = plt.figure('Network Optimizer', figsize=(16, 10))

    if len(bounds) == 2:
        width_ratios = [2.15, 2.15, 0.9]
    else:
        width_ratios = [2.15, 2.15, 0.75]
    spec = gridspec.GridSpec(ncols=3, nrows=2,
                             width_ratios=width_ratios)  # , hspace=0.075)

    mng = plt.get_current_fig_manager()
    mng.full_screen_toggle()

    start_time = time.time()
    best = None
    flag = True

    def s_mag(goto):
        global start_time, positions, best, ntw_best, s, grid_x, \
            grid_y, s11_progress, s21_progress, f_goal_label, flag
        try:

            # Float to Int

            goto = [int(pos) for pos in goto]

            if not flag:
                positions = np.append(positions, [goto], axis=0)
            with open('log.txt', 'a') as log:
                log.write('==================================\n')
                for parameter_idx in range(len(parameters)):
                    log.write('[*] Shifting x' + str(parameter_idx + 1)
                              + ' to position: '
                              + str(goto[parameter_idx]) + '\n')

            # print('a')

            threads = []
            plt.tight_layout(pad=0.7)
            plt.show()
            plt.draw()
            plt.pause(0.01)
            plt.clf()

            # for plt_thread in plt_threads:
            # kil....threads.append(threading.Thread(target=plt_thread, args=(0,)))

            for parameter_idx in range(len(parameters)):
                threads.append(threading.Thread(target=modules[parameter_idx].drive,
                               args=(goto[parameter_idx], )))
            for thread in threads:
                thread.start()
            for thread in threads:
                thread.join()

            # Measure current S11 and S21

            s11 = nv.data(0)
            s21 = nv.data(1)

            f_arr = np.linspace(f_start / 1e6, f_stop / 1e6,
                                num=sweep_pts)
            f = skrf.Frequency.from_f(f_arr, unit='MHz')
            s = np.zeros((len(f), 2, 2), dtype='complex_')
            s[:, 0, 0] = s11
            s[:, 0, 1] = s21  # # np.ones_like(s11)*np.nan #np.zeros_like(s11) # S12 (incompatible with NanoVNA)
            s[:, 1, 0] = s21
            s[:, 1, 1] = s11  # # np.ones_like(s11)*np.nan #np.ones_like(s11) # S22 (incompatible with NanoVNA)

            ntw = skrf.Network(frequency=f, s=s)
            if calibrated:
                ntw = calibration.apply_cal((ntw, ntw))

            s = ntw.s
            if not flag:
                s11_progress = np.append(s11_progress,
                        skrf.mathFunctions.complex_2_db(s[idx, 0, 0]))
                s21_progress = np.append(s21_progress,
                        skrf.mathFunctions.complex_2_db(s[idx, 1, 0]))

            ntw.write_touchstone('Current', skrf_comment=False)
            ntw = skrf.Network('Current.s2p')

            if flag:
                best = sign * skrf.mathFunctions.complex_2_db(s[idx, 1,
                        0])
                ntw_best = ntw
                ntw_best.write_touchstone('Best', skrf_comment=False)
                ntw_best = skrf.Network('Best.s2p')
            elif sign * skrf.mathFunctions.complex_2_db(s[idx, 1, 0]) \
                < best:
                best = sign * skrf.mathFunctions.complex_2_db(s[idx, 1,
                        0])
                ntw_best = ntw
                ntw_best.write_touchstone('Best', skrf_comment=False)
                ntw_best = skrf.Network('Best.s2p')

            ax1 = fig.add_subplot(spec[0])  # Initial vs Best
            ax2 = fig.add_subplot(spec[3])  # Current
            ax3 = fig.add_subplot(spec[4])  # Positions vs Iteration
            ax4 = fig.add_subplot(spec[1])  # |S11|, |S21| vs Iteration
            ax5 = fig.add_subplot(spec[1:, -1])  # x1 vs x2 (S21)
            ax6 = fig.add_subplot(spec[2])  # x1 vs x2 (S11)
            if len(bounds) != 2:
                ax5.axis('off')
                ax6.axis('off')
            ax3.xaxis.set_major_locator(MaxNLocator(integer=True))  # Force integer x-ticks
            ax4.xaxis.set_major_locator(MaxNLocator(integer=True))  # Force integer x-ticks

            # plt.subplots_adjust(wspace=0,hspace=0) #hspace=0.075)

            # Initial vs Best

            ax1.set_title('$\mathrm{Optimization \ Progress}$',
                          fontsize=19)

            ax1.axvline(x=f_goal * 1e6, color='#9467bd', linestyle='--')
            ntw_init.plot_s_db(n=0, m=0, ax=ax1)
            ntw_init.plot_s_db(n=0, m=1, ax=ax1)

            ntw_best.plot_s_db(n=0, m=0, ax=ax1)
            ntw_best.plot_s_db(n=0, m=1, ax=ax1)

            ax1.set_xlabel('$\mathrm{Frequency \ (MHz)}$')
            ax1.set_ylabel('$\mathrm{Magnitude \ (dB)}$')
            ax1.set_ylim(top=0)
            ax1.grid()
            ax1.legend(loc='lower left')

            # Current

            ax2.axvline(x=f_goal * 1e6, color='#9467bd', linestyle='--')
            ntw.plot_s_db(n=0, m=0, ax=ax2)
            ntw.plot_s_db(n=0, m=1, ax=ax2)
            ax1.get_shared_x_axes().join(ax1, ax2)
            ax1.set_xticklabels([])
            x_axis = ax1.axes.get_xaxis()
            x_axis.set_label_text('')
            x_label = x_axis.get_label()
            x_label.set_visible(False)
            ax2.set_ylabel('$\mathrm{Magnitude \ (dB)}$')
            ax2.set_ylim(top=0)
            ax2.grid()
            ax2.legend(loc='lower left')

            # Positions vs Iteration #

            for parameter_idx in range(len(parameters)):
                ax3.plot(range(1, len(positions) + 1), positions[:,
                         parameter_idx], label='$x_{'
                         + str(parameter_idx + 1) + r'}$')
            ax3.set_xlim(left=1)
            ax3.grid()
            if len(positions) < 2:
                ax3.set_xticklabels([])
            ax3.legend(loc='best')
            ax3.set_xlabel('$\mathrm{Iteration \ \#}$')
            ax3.set_ylabel('$\mathrm{Position \ (steps)}$')

            # |S11|, |S21| vs Iteration #

            ax4.set_title('$\mathrm{Parameter \ Sequence}$',
                          fontsize=19)
            ax4.plot(range(1, len(s11_progress) + 1), s11_progress,
                     label='$S_{11} \ \mathrm{(' + f_goal_label
                     + ' \ \mathrm{MHz})}$')
            ax4.plot(range(1, len(s21_progress) + 1), s21_progress,
                     label='$S_{21} \ \mathrm{(' + f_goal_label
                     + ' \ \mathrm{MHz})}$')
            ax4.set_xlim(left=1)
            ax4.get_shared_x_axes().join(ax4, ax3)
            ax4.set_xticklabels([])
            x_axis = ax1.axes.get_xaxis()
            x_axis.set_label_text('')
            x_label = x_axis.get_label()
            x_label.set_visible(False)
            ax4.grid()
            ax4.legend(loc='best')
            ax4.set_ylabel('$\mathrm{Magnitude \ (dB)}$')

            # |S21| vs x1 vs x2

            if len(bounds) == 2:
                ax5.set_aspect('equal', anchor='N')
                if not flag and len(positions) >= 4:
                    grid_z2 = interpolate.griddata(positions,
                            s21_progress, (grid_x, grid_y),
                            method='cubic')
                    ax5.imshow(grid_z2.T, extent=(bounds[0][0],
                               bounds[0][1], bounds[1][0],
                               bounds[1][1]), origin='lower')
                    levels = np.arange(np.amin(s21_progress),
                            np.amax(s21_progress),
                            abs(np.amax(s21_progress)
                            - np.amin(s21_progress)) / 3)
                    ax5.contour(grid_z2.T, levels, colors='k',
                                origin='lower', extent=(bounds[0][0],
                                bounds[0][1], bounds[1][0],
                                bounds[1][1]))
                    ax5.plot(positions[:, 0], positions[:, 1], 'r.',
                             ms=3)
                    ax5.set_aspect((bounds[0][1] - bounds[0][0])
                                   / (bounds[1][1] - bounds[1][0]),
                                   anchor='N')
                ax5.set_xlabel('$x_{1}$')
                ax6.set_xticks([])
                ax6.get_shared_x_axes().join(ax6, ax5)
                ax6.set_xticklabels([])
                x_axis = ax6.axes.get_xaxis()
                x_axis.set_label_text('')
                x_label = x_axis.get_label()
                x_label.set_visible(False)
                ax5.set_ylabel('$x_{2}$')

                # |S11| vs x1 vs x2

                ax6.set_aspect('equal', anchor='S')
                if not flag and len(positions) >= 4:
                    grid_z2 = interpolate.griddata(positions,
                            s11_progress, (grid_x, grid_y),
                            method='cubic')
                    ax6.imshow(grid_z2.T, extent=(bounds[0][0],
                               bounds[0][1], bounds[1][0],
                               bounds[1][1]), origin='lower')
                    levels = np.arange(np.amin(s11_progress),
                            np.amax(s11_progress),
                            abs(np.amax(s11_progress)
                            - np.amin(s11_progress)) / 3)
                    ax6.contour(grid_z2.T, levels, colors='k',
                                origin='lower', extent=(bounds[0][0],
                                bounds[0][1], bounds[1][0],
                                bounds[1][1]))
                    ax6.plot(positions[:, 0], positions[:, 1], 'r.',
                             ms=3)
                    ax6.set_aspect((bounds[0][1] - bounds[0][0])
                                   / (bounds[1][1] - bounds[1][0]),
                                   anchor='S')

                ax5.set_xlabel('$x_{1}$')
                ax6.set_title('$\mathrm{Evaluation \ Heatmaps}$\n',
                              fontsize=19)
                ax6.set_xlabel('$x_{1}$')
                ax6.set_ylabel('$x_{2}$')

                fig.text(x=0.91, y=0.786, s=r'$|S_{11}|$', fontsize=13)
                fig.text(x=0.91, y=0.506, s=r'$|S_{21}|$', fontsize=13)

            fig.text(x=0.850, y=0.926,
                     s=r'$\mathrm{\underline{Optimize\textbf{RF}}}$',
                     fontsize=26, color='seagreen')

            y_offset = 0
            if len(bounds) != 2:
                y_offset = 0.6

            fig.text(x=0.847, y=0.160 + y_offset,
                     s=r'$\mathrm{\underline{Optimization \ Status}}$',
                     fontsize=16)

            if len(positions) >= int(exp_ratio * max_calls):
                search_space = 'Local \ (exploiting)'
            else:
                search_space = 'Global \ (exploring)'

            fig.text(x=0.820, y=0.125 + y_offset,
                     s=r'$\mathrm{Search \ Space}$: $\mathrm{'
                     + search_space + r'}$', fontsize=14)

            # Check convergence

            convergence = 'Not \ reached'
            if s_parameter.lower() == 's21':
                s_progress = s21_progress
            else:
                s_progress = s11_progress
            if len(s_progress) > int(exp_ratio * max_calls):
                s_progress_reversed = s_progress[::-1]
                if sign == -1:  # Maximize magnitude
                    last_best_idx = np.argmax(s_progress_reversed)
                else:

                      # Minimize magnitude

                    last_best_idx = np.argmin(s_progress_reversed)
                if last_best_idx >= int(conv_ratio * max_calls):
                    convergence = 'Reached'

            fig.text(x=0.839, y=0.095 + y_offset,
                     s=r'$\mathrm{Convergence}$: $\mathrm{'
                     + convergence + r'}$', fontsize=14)

            percentage = str(round(100 * (len(positions) / max_calls),
                             1))
            fig.text(x=0.844, y=0.065 + y_offset,
                     s=r'$\mathrm{Progress}$: $\mathrm{'
                     + str(len(positions)) + r'/' + str(max_calls)
                     + r'\ (' + percentage + r'\%)}$', fontsize=14)

            runtime = int(time.time() - start_time)
            fig.text(x=0.866, y=0.035 + y_offset,
                     s=r'$\mathrm{Runtime}$: $\mathrm{'
                     + str(timedelta(seconds=runtime))[:-3].replace(':'
                     , r'}$:$\mathrm{') + r'}$', fontsize=14)

            if flag:
                flag = False
            plt.savefig(str(len(positions) + 1) + '.png', dpi=300)
            if len(positions) == max_calls and output != '':
                plt.savefig(output, dpi=300)
        except KeyboardInterrupt:

            # plt.show()
            # plt.draw()
            # plt.pause(0.01)
            # plt.clf()

            raise ValueError('\n[-] Sweep terminated by user.')

        if s_parameter.lower() == 's21':
            magnitude = sign * skrf.mathFunctions.complex_2_db(s[idx,
                    1, 0])
        else:
            magnitude = sign * skrf.mathFunctions.complex_2_db(s[idx,
                    0, 0])

        return magnitude

    # ## Optimization ###

    try:

        # Initiate plot (goto=0 to begin plot)

        with open('log.txt', 'a') as log:
            log.write('=-=-=-=-=-=-=-=-=-=-=-=-=-=-=\n')
            log.write('[+] Initiating parameters...\n')
        s_mag([0] * len(bounds))

        # Begin optimization

        result = gp_minimize(
            s_mag,
            bounds,
            n_calls=max_calls,
            n_initial_points=int(exp_ratio * max_calls),
            initial_point_generator='lhs',
            verbose=True,
            )
    except Exception, e:

        with open('log.txt', 'a') as log:
            print e
            log.write(str(e))
            log.write('''
[-] Optimization interrupted.
''')

    if reset:
        with open('log.txt', 'a') as log:
            if len(bounds) == 1:
                log.write('[*] Resetting parameter back to initial position (0 steps).\n'
                          )
            else:
                log.write('[*] Resetting parameters back to initial positions (0 steps).\n'
                          )

        threads = []
        for parameter_idx in range(len(parameters)):
            threads.append(threading.Thread(target=modules[parameter_idx].drive,
                           args=(0, )))

        for thread in threads:
            thread.start()
        for thread in threads:
            thread.join()

        for parameter_idx in range(len(parameters)):
            modules[parameter_idx].cleanup()


def main(gui=False):

    # Load argument values

    parser = argparse.ArgumentParser()

    if gui:

        # GUI

        parser.add_argument('-G', '--gui', dest='gui',
                            action='store_true',
                            help='Start graphical user interface')
        parser.set_defaults(gui=False)
    else:

        # GUI (included for -h)

        parser.add_argument('-G', '--gui', dest='gui',
                            action='store_true',
                            help='Start graphical user interface')
        parser.set_defaults(gui=False)

        # Sweep arguments

        parser.add_argument(
            '-f0',
            '--f_start',
            dest='f_start',
            help='Start frequency (Hz)',
            type=float,
            required=True,
            )
        parser.add_argument(
            '-f1',
            '--f_stop',
            dest='f_stop',
            help='Stop frequency (Hz)',
            type=float,
            required=True,
            )
        parser.add_argument(
            '-t',
            '--sweep_pts',
            dest='sweep_pts',
            help='Sweep points (Hz)',
            type=int,
            default=101,
            )

        # Measurement calibration

        parser.add_argument('-c', '--cal', dest='cal',
                            action='store_true',
                            help='Perform interactive measurement calibration'
                            )
        parser.set_defaults(cal=False)
        parser.add_argument(
            '-l',
            '--load_cal',
            dest='load_cal',
            help='Import local measurement calibration from directory',
            type=str,
            default='',
            )

        # Parameter properties

        parser.add_argument(
            '-p',
            '--parameters',
            dest='parameters',
            nargs='+',
            help='Tuning parameters (module names)',
            type=str,
            required=True,
            )
        parser.add_argument(
            '-b',
            '--bounds',
            dest='bounds',
            help='Parameters boundaries. Format: \'[(x1_min, x1_max), (x2_min, x2_max), ..., (xn_min, xn_max)]\''
                ,
            type=str,
            required=True,
            )
        parser.add_argument('-r', '--reset', dest='reset',
                            action='store_true',
                            help='Reset parameters back to initial positions'
                            )
        parser.set_defaults(reset=True)

        # Goal arguments

        parser.add_argument(
            '-f',
            '--f_goal',
            dest='f_goal',
            help='Target frequency (Hz)',
            type=float,
            required=True,
            )
        parser.add_argument(
            '-g',
            '--goal',
            dest='goal',
            help='Goal operator (minimize/maximize)',
            type=str,
            default='max',
            )
        parser.add_argument(
            '-s',
            '--s_parameter',
            dest='s_parameter',
            help='Scattering parameter\'s magnitude to optimize',
            type=str,
            default='S21',
            )

        # Optimizer settings

        parser.add_argument(
            '-i',
            '--max_calls',
            dest='max_calls',
            help='Maximum number of function iterations (calls)',
            type=int,
            default=100,
            )
        parser.add_argument(
            '-e',
            '--exp_ratio',
            dest='exp_ratio',
            help='Exploration ratio (initial global search)',
            type=float,
            default=0.1,
            )
        parser.add_argument(
            '-q',
            '--conv_ratio',
            dest='conv_ratio',
            help='Convergence ratio',
            type=float,
            default=0.3,
            )

        # Result output

        parser.add_argument(
            '-o',
            '--output',
            dest='output',
            help='Results output filename',
            type=str,
            default='',
            )
    args = parser.parse_args()

    if not gui:

        # ## Begin optimization ###

        optimize(
            args.f_start,
            args.f_stop,
            args.sweep_pts,
            args.cal,
            args.load_cal,
            args.f_goal,
            args.goal,
            args.s_parameter,
            args.parameters,
            args.bounds,
            args.reset,
            args.max_calls,
            args.exp_ratio,
            args.conv_ratio,
            args.output,
            )
    else:

        # Create window object

        app = Tk()

        def start_optimizer():
            top = Toplevel()
            top.geometry('500x100')
            top.title('OptimizeRF')
            Message(top, text='Starting Optimizer...', font=('Helvetica'
                    , 18, 'italic'), padx=50, pady=50).pack()
            top.after(10000, top.destroy)
            time.sleep(1)

            # ## Begin optimization ###
            # print(
            # ....float(f_start_text.get()),
            # ....float(f_stop_text.get()),
            # ....int(sweep_pts_text.get()),
            # ....var_cal.get() == 1,
            # ....str(load_cal_entry.get()) if str(load_cal_entry.get()) != '' else '',
            # ....float(f_goal_text.get()),
            # ....str((goal_text.get()).lower()),
            # ....str(objective),
            # ....list(str(parameters_text.get()).split()),
            # ....str(bounds_text.get()),
            # ....int(var_reset.get()) == 1,
            # ....int(max_calls_text.get()),
            # ....float(exploration_ratio_text.get()),
            # ....float(convergence_ratio_text.get()),
            # ....str(output_text.get())
            # )

            optimize(
                float(f_start_text.get()),
                float(f_stop_text.get()),
                int(sweep_pts_text.get()),
                var_cal.get() == 1,
                (str(load_cal_entry.get()) if str(load_cal_entry.get())
                 != '' else ''),
                float(f_goal_text.get()),
                str(goal_text.get().lower()),
                str(objective),
                list(str(parameters_text.get()).split()),
                str(bounds_text.get()),
                int(var_reset.get()) == 1,
                int(max_calls_text.get()),
                float(exploration_ratio_text.get()) / 100,
                float(convergence_ratio_text.get()) / 100,
                str(output_text.get()),
                )

        # ## Sweep arguments ###

        sweep_args_label = Label(app, text='Sweep Arguments',
                                 font=('Helvetica', 18, 'bold',
                                 'underline'))
        sweep_args_label.grid(
            row=0,
            column=0,
            columnspan=2,
            pady=10,
            padx=10,
            sticky=W,
            )

        # f_start

        f_start_text = DoubleVar()
        f_start_label = Label(app, text=' Start frequency: ',
                              font=('Helvetica', 14))
        f_start_label.grid(row=1, column=0, padx=20, sticky=W)
        f_start_entry = Entry(app, textvariable=f_start_text, width=12)
        f_start_entry.grid(row=1, column=1)
        f_start_label_hz = Label(app, text='Hz', font=('Helvetica', 14))
        f_start_label_hz.grid(row=1, column=2, sticky=W)

        # f_stop

        f_stop_text = DoubleVar()
        f_stop_label = Label(app, text=' Stop frequency: ',
                             font=('Helvetica', 14))
        f_stop_label.grid(row=2, column=0, padx=20, sticky=W)
        f_stop_entry = Entry(app, textvariable=f_stop_text, width=12)
        f_stop_entry.grid(row=2, column=1)
        f_stop_label_hz = Label(app, text='Hz', font=('Helvetica', 14))
        f_stop_label_hz.grid(row=2, column=2, sticky=W)

        # sweep_pts

        sweep_pts_text = IntVar(value=101)
        sweep_pts_label = Label(app, text=' Sweep Points: ',
                                font=('Helvetica', 14))
        sweep_pts_label.grid(row=3, column=0, padx=20, sticky=W)
        sweep_pts_entry = Entry(app, textvariable=sweep_pts_text,
                                width=12)
        sweep_pts_entry.grid(row=3, column=1)

        # ## Measurement Calibration ###

        measurement_calibration_label = Label(app,
                text='Measurement Calibration', font=('Helvetica', 18,
                'bold', 'underline'))
        measurement_calibration_label.grid(
            row=4,
            column=0,
            columnspan=3,
            pady=10,
            padx=10,
            sticky=W,
            )
        var_cal = IntVar()

        # cal

        cal = Radiobutton(app, text='Perform interactive calibration',
                          variable=var_cal, value=1, font=('Helvetica',
                          14))
        cal.grid(row=5, column=0, columnspan=3, padx=10, sticky=W)

        # load_cal

        load_cal_text = StringVar(value='measured')
        load_cal = Radiobutton(app,
                               text='Import local calibration from dir:'
                               , variable=var_cal, value=2,
                               font=('Helvetica', 14))
        load_cal.grid(row=6, column=0, columnspan=3, padx=10, sticky=W)
        load_cal_entry = Entry(app, textvariable=load_cal_text, width=9)
        load_cal_entry.grid(row=6, column=1, padx=30)

        # uncalibrated

        uncalibrated = Radiobutton(app, text='None (uncalibrated)',
                                   variable=var_cal, value=3,
                                   font=('Helvetica', 14))
        uncalibrated.grid(row=7, column=0, columnspan=3, padx=10,
                          sticky=W)
        uncalibrated.select()

        # ## Parameters ###

        params_label = Label(app, text='Parameter Properties',
                             font=('Helvetica', 18, 'bold', 'underline'
                             ))
        params_label.grid(
            row=8,
            column=0,
            columnspan=3,
            pady=10,
            padx=10,
            sticky=W,
            )

        # parameters

        parameters_text = StringVar(value='module_x1 module_x2')
        parameters_label = Label(app, text=' Parameters:',
                                 font=('Helvetica', 14))
        parameters_label.grid(row=9, column=0, padx=20, sticky=W)
        parameters_entry = Entry(app, textvariable=parameters_text,
                                 width=32)
        parameters_entry.grid(row=9, column=1, columnspan=3)

        # boundaries

        bounds_text = \
            StringVar(value='[(x1_min, x1_max), (x2_min, x2_max)]')
        bounds_label = Label(app, text=' Boundaries:', font=('Helvetica'
                             , 14))
        bounds_label.grid(row=10, column=0, padx=20, sticky=W)
        bounds_entry = Entry(app, textvariable=bounds_text, width=32)
        bounds_entry.grid(row=10, column=1, columnspan=3)

        # reset

        var_reset = IntVar()
        reset = Checkbutton(app,
                            text='Reset parameters back to initial positions'
                            , variable=var_reset, font=('Helvetica',
                            14))
        reset.grid(row=11, column=0, columnspan=11, padx=10, sticky=W)
        reset.select()

        # ## Goals ###

        goals_label = Label(app, text='Optimization Goal',
                            font=('Helvetica', 18, 'bold', 'underline'))
        goals_label.grid(
            row=12,
            column=0,
            columnspan=3,
            pady=10,
            padx=10,
            sticky=W,
            )

        # f_goal

        f_goal_text = DoubleVar()
        f_goal_label = Label(app, text=' Target frequency: ',
                             font=('Helvetica', 14))
        f_goal_label.grid(row=13, column=0, padx=20, sticky=W)
        f_goal_entry = Entry(app, textvariable=f_goal_text, width=12)
        f_goal_entry.grid(row=13, column=1)
        f_goal_label_hz = Label(app, text='Hz', font=('Helvetica', 14))
        f_goal_label_hz.grid(row=13, column=2, sticky=W)

        # goal

        goal_text = StringVar(app)
        goal_text.set('Maximize')
        goal_label = Label(app, text=' Optimization goal:',
                           font=('Helvetica', 14))
        goal_label.grid(row=14, column=0, padx=20, sticky=W)
        goal_entry = OptionMenu(app, goal_text, 'Minimize', 'Maximize')
        goal_entry.grid(row=14, column=1)

        # objective S-Parameter

        objective_text = StringVar(app)
        objective_text.set('     |S21|  ')
        objective_label = Label(app, text=' Objective S-Parameter:',
                                font=('Helvetica', 14))
        objective_label.grid(row=15, column=0, padx=20, sticky=W)
        objective_entry = OptionMenu(app, objective_text, '     |S11|  '
                , '     |S21|  ')
        objective_entry.grid(row=15, column=1)
        if objective_text.get() == '     |S21|  ':
            objective = 'S21'
        else:
            objective = 'S11'

        # ## Optimizer Settings ###

        settings_label = Label(app, text='Optimizer Settings',
                               font=('Helvetica', 18, 'bold',
                               'underline'))
        settings_label.grid(
            row=16,
            column=0,
            columnspan=3,
            pady=10,
            padx=10,
            sticky=W,
            )

        # max_calls

        max_calls_text = IntVar(value=100)
        max_calls_label = Label(app, text=' Max calls: ',
                                font=('Helvetica', 14))
        max_calls_label.grid(row=17, column=0, padx=20, sticky=W)
        max_calls_entry = Entry(app, textvariable=max_calls_text,
                                width=12)
        max_calls_entry.grid(row=17, column=1)

        # exploration_ratio

        exploration_ratio_text = IntVar(value=15)
        exploration_ratio_label = Label(app, text=' Exploration ratio: '
                , font=('Helvetica', 14))
        exploration_ratio_label.grid(row=18, column=0, padx=20,
                sticky=W)
        exploration_ratio_entry = Entry(app,
                textvariable=exploration_ratio_text, width=12)
        exploration_ratio_entry.grid(row=18, column=1)
        exploration_ratio_label_percent = Label(app, text='%',
                font=('Helvetica', 14))
        exploration_ratio_label_percent.grid(row=18, column=2, sticky=W)

        # convergence_ratio

        convergence_ratio_text = IntVar(value=25)
        convergence_ratio_label = Label(app, text=' Convergence ratio: '
                , font=('Helvetica', 14))
        convergence_ratio_label.grid(row=19, column=0, padx=20,
                sticky=W)
        convergence_ratio_entry = Entry(app,
                textvariable=convergence_ratio_text, width=12)
        convergence_ratio_entry.grid(row=19, column=1)
        convergence_ratio_label_percent = Label(app, text='%',
                font=('Helvetica', 14))
        convergence_ratio_label_percent.grid(row=19, column=2, sticky=W)

        # ## Results ###

        results_label = Label(app, text='Results', font=('Helvetica',
                              18, 'bold', 'underline'))
        results_label.grid(
            row=20,
            column=0,
            columnspan=3,
            pady=10,
            padx=10,
            sticky=W,
            )

        # output

        output_text = StringVar(value='results.pdf')
        output_label = Label(app, text=' Output filename:',
                             font=('Helvetica', 14))
        output_label.grid(row=21, column=0, padx=20, sticky=W)
        output_entry = Entry(app, textvariable=output_text, width=12)
        output_entry.grid(row=21, column=1)

        # fullscreen

        var_fullscreen = IntVar()
        fullscreen = Checkbutton(app, text='Enable fullscreen',
                                 variable=var_fullscreen,
                                 font=('Helvetica', 14))
        fullscreen.grid(row=22, column=0, columnspan=11, padx=10,
                        sticky=W)
        fullscreen.select()
        start = Button(app, text='Start Optimizer',
                       command=start_optimizer, width=59, bg='#2edc71')
        start.grid(
            row=23,
            column=0,
            columnspan=3,
            padx=10,
            sticky=W,
            pady=10,
            )

        app.title('OptimizeRF | GUI')
        app.geometry('520x960')

        # Start GUI

        app.mainloop()


if __name__ == '__main__':
    if '-G' in sys.argv or '--gui' in sys.argv:
        gui = True
    else:
        gui = False
    main(gui)
\end{lstlisting}

\nomenclature{VNA}{Vector Network Analyzer}


\end{document}